\documentclass[modern]{aastex62}

\usepackage{newtxtext,newtxmath}

\usepackage[T1]{fontenc}
\usepackage{ae,aecompl}
\usepackage{longtable}
\usepackage{ulem}
\usepackage{newunicodechar,graphicx}

\DeclareRobustCommand{\okina}{%
  \raisebox{\dimexpr\fontcharht\font`A-\height}{%
    \scalebox{0.8}{`}%
  }%
}
\newunicodechar{ʻ}{\okina}

\usepackage{graphicx}	
\usepackage{amsmath}	
\usepackage{amssymb}	
\usepackage{xspace}
\usepackage{fontawesome}

\newcommand{\numax}{\mbox{$\nu_{\rm max}$}\xspace}
\newcommand{\Dnu}{\mbox{$\Delta \nu$}\xspace}
\newcommand{\dnu}{\mbox{$\delta \nu$}\xspace}
\newcommand{\muHz}{\mbox{$\mu$Hz}\xspace}
\newcommand{\teff}{\mbox{$T_{\rm eff}$}\xspace}
\newcommand{\logg}{\mbox{$\log g$}\xspace}
\newcommand{\feh}{\mbox{$\rm{[Fe/H]}$}\xspace}
\newcommand{\msun}{\mbox{$\mathrm{M}_{\sun}$}\xspace}
\newcommand{\rsun}{\mbox{$\mathrm{R}_{\sun}$}\xspace}
\newcommand{\kepler}{\textit{Kepler}\xspace}
\newcommand{\ktwo}{\textit{K2}\xspace}
\newcommand{\hipparcos}{\textit{Hipparcos}\xspace}
\newcommand{\gaia}{\textit{Gaia}\xspace}
\definecolor{linkcolor}{rgb}{0.1216,0.4667,0.7059}

\usepackage{xcolor, fontawesome}
\definecolor{twitterblue}{RGB}{64,153,255}
\newcommand\twitter[1]{\href{https://twitter.com/#1 }{\textcolor{twitterblue}{\faTwitter}\,\tt \textcolor{twitterblue}{@#1}}}

\submitjournal{ApJ}

\begin{document}

\title{The \textit{Kepler} Smear Campaign: Light curves for 102 Very Bright Stars}
\shorttitle{The Kepler Smear Campaign}
\shortauthors{B. J. S. Pope et al.}

\correspondingauthor{Benjamin J. S. Pope \twitter{fringetracker}}
\email{benjamin.pope@nyu.edu}

\author[0000-0003-2595-9114]{Benjamin J. S. Pope}
\affiliation{Center for Cosmology and Particle Physics, Department of Physics, New York University, 726 Broadway, New York, NY 10003, USA}
\affiliation{Center for Data Science, New York University, 60 Fifth Ave, New York, NY 10011, USA}
\affiliation{NASA Sagan Fellow}
\affiliation{Oxford Astrophysics, Denys Wilkinson Building, University of Oxford, OX1 3RH, Oxford, UK}

\author[0000-0002-4290-7351]{Guy R. Davies}
\affiliation{School of Physics and Astronomy, University of Birmingham, Birmingham B15 2TT, UK}
\affiliation{Stellar Astrophysics Centre, Department of Physics and Astronomy, Aarhus University, Ny Munkegade 120, DK-8000 Aarhus C, Denmark}

\author[0000-0002-1423-2174]{Keith Hawkins}
\affiliation{Department of Astronomy, The University of Texas at Austin, 2515 Speedway Boulevard, Austin, TX 78712, USA}

\author[0000-0002-6980-3392]{Timothy R. White}
\affiliation{Research School of Astronomy and Astrophysics, Mount Stromlo Observatory, The Australian National University, Canberra, ACT 2611, Australia}
\affiliation{Stellar Astrophysics Centre, Department of Physics and Astronomy, Aarhus University, Ny Munkegade 120, DK-8000 Aarhus C, Denmark}

\author[0000-0002-5496-365X]{Amalie Stokholm}
\affiliation{Stellar Astrophysics Centre, Department of Physics and Astronomy, Aarhus University, Ny Munkegade 120, DK-8000 Aarhus C, Denmark}

\author[0000-0001-6637-5401]{Allyson Bieryla}
\affiliation{Center for Astrophysics | Harvard \& Smithsonian, 60 Garden Street, Cambridge, MA 02138, USA}

\author[0000-0001-9911-7388]{David W. Latham}
\affiliation{Center for Astrophysics | Harvard \& Smithsonian, 60 Garden Street, Cambridge, MA 02138, USA}

\author[0000-0001-7297-8508]{Madeline Lucey}
\affiliation{Department of Astronomy, The University of Texas at Austin, 2515 Speedway Boulevard, Austin, TX 78712, USA}

\author[0000-0003-1822-7126]{Conny Aerts}
\affiliation{Instituut voor Sterrenkunde, KU Leuven, Celestijnenlaan 200D, B-3001 Leuven, Belgium}
\affiliation{Department of Astrophysics, IMAPP, Radboud University Nijmegen, P.O. Box 9010, NL-6500 GL Nijmegen, The Netherlands}

\author[0000-0003-1453-0574]{Suzanne Aigrain}
\affiliation{Oxford Astrophysics, Denys Wilkinson Building, University of Oxford, OX1 3RH, Oxford, UK}

\author[0000-0002-0865-3650]{Victoria Antoci}
\affiliation{Stellar Astrophysics Centre, Department of Physics and Astronomy, Aarhus University, Ny Munkegade 120, DK-8000 Aarhus C, Denmark}

\author[0000-0001-5222-4661]{Timothy R. Bedding}
\affiliation{Sydney Institute for Astronomy, School of Physics A28, The University of Sydney, NSW 2006, Australia}
\affiliation{Stellar Astrophysics Centre, Department of Physics and Astronomy, Aarhus University, DK-8000 Aarhus C, Denmark}

\author[0000-0001-7402-3852]{Dominic M. Bowman}
\affiliation{Instituut voor Sterrenkunde, KU Leuven, Celestijnenlaan 200D, B-3001 Leuven, Belgium}

\author[0000-0003-1963-9616]{Douglas A. Caldwell}
\affiliation{SETI Institute, 189 Bernardo Avenue, Mountain View, CA 94043, USA}

\author[0000-0003-1125-2564	]{Ashley Chontos}
\affiliation{Institute for Astronomy, University of Hawai\okina i, 2680 Woodlawn Drive, Honolulu, HI 96822, USA}

\author[0000-0002-9789-5474]{Gilbert A. Esquerdo}
\affiliation{Harvard-Smithsonian Center for Astrophysics, 60 Garden Street, Cambridge, MA 02138, USA}

\author[0000-0001-8832-4488 ]{Daniel Huber}
\affiliation{Institute for Astronomy, University of Hawai\okina i, 2680 Woodlawn Drive, Honolulu, HI 96822, USA}
\affiliation{SETI Institute, 189 Bernardo Avenue, Mountain View, CA 94043, USA}
\affiliation{Stellar Astrophysics Centre, Department of Physics and Astronomy, Aarhus University, DK-8000 Aarhus C, Denmark}

\author[0000-0002-0722-7406]{Paula Jofr\'{e}}
\affiliation{N\'{u}cleo de Astronom\'{i}a, Facultad de Ingenier\'{i}a y Ciencias, Universidad Diego Portales, Ej\'{e}rcito 441, Santiago De, Chile}

\author[0000-0002-5648-3107]{Simon J. Murphy}
\affiliation{Sydney Institute for Astronomy, School of Physics A28, The University of Sydney, NSW 2006, Australia}
\affiliation{Stellar Astrophysics Centre, Department of Physics and Astronomy, Aarhus University, DK-8000 Aarhus C, Denmark}

\author[0000-0003-2771-1745]{Timothy van Reeth}
\affiliation{Sydney Institute for Astronomy, School of Physics A28, The University of Sydney, NSW 2006, Australia}
\affiliation{Stellar Astrophysics Centre, Department of Physics and Astronomy, Aarhus University, DK-8000 Aarhus C, Denmark}

\author[0000-0002-6137-903X]{Victor Silva Aguirre}
\affiliation{Stellar Astrophysics Centre, Department of Physics and Astronomy, Aarhus University, DK-8000 Aarhus C, Denmark}

\author[0000-0002-0007-6211]{Jie Yu}
\affiliation{Sydney Institute for Astronomy, School of Physics A28, The University of Sydney, NSW 2006, Australia}
\affiliation{Stellar Astrophysics Centre, Department of Physics and Astronomy, Aarhus University, DK-8000 Aarhus C, Denmark}


\begin{abstract}
We present the first data release of the \kepler Smear Campaign, using collateral `smear' data obtained in the \kepler four-year mission to reconstruct light curves of 102~stars too bright to have been otherwise targeted. We describe the pipeline developed to extract and calibrate these light curves, and show that we attain photometric precision comparable to stars analyzed by the standard pipeline in the nominal \kepler mission. In this paper, aside from publishing the light curves of these stars, we focus on 66~red giants for which we detect solar-like oscillations, characterizing 33 of these in detail with spectroscopic chemical abundances and asteroseismic masses as benchmark stars. We also classify the whole sample, finding nearly all to be variable, with classical pulsations and binary effects. All source code, light curves, TRES spectra, and asteroseismic and stellar parameters are publicly available as a \kepler legacy sample. \href{https://github.com/benjaminpope/smearcampaign}{\color{linkcolor}\faGithub}
\end{abstract}




\section{Introduction}
\label{intro}


\kepler has revolutionized the field of asteroseismology both for solar-like oscillations \citep{2010PASP..122..131G,2010ApJ...713L.169C} and for coherent heat-engine driven oscillations \citep{2018arXiv180907779A}. It has yielded the detection of gravity-dominated mixed-mode period spacings for red giants \citep{rggmodes,2014A&A...572L...5M}, enabling probes of interior rotation \citep{rggmoderotation,2012A&A...548A..10M,2012ApJ...756...19D} and distinguishing between hydrogen- and helium-burning cores \citep{rggmodehelium,2012A&A...540A.143M}. It has also permitted the determination of ages and fundamental parameters of cool main-sequence stars \citep{silvaages}, including planet-hosting stars \citep{huberplanetages,silvaaguirre2015,2016MNRAS.456.2183D,2018MNRAS.479.4786V}. \kepler gravity-mode asteroseismology has also been used to derive the internal rotation profiles of intermediate mass stars \citep{triana15,vanreeth18}. 

A major outcome of the \kepler asteroseismology programme is a legacy sample of extremely well-characterized stars that can serve as benchmarks for future work \citep{keplerlegacy1,silvaaguirre2017,silvaaguirre2015,2016MNRAS.456.2183D}. Asteroseismological studies with \kepler complement other probes of stellar physics, such as the APOKASC sample of~1916 spectroscopically- and asteroseismically-characterized red giant stars \citep{2014ApJS..215...19P}. For this APOKASC sample, \citet{hawkinsapogee} have been able to extract precise elemental abundances by fitting spectroscopic data with \logg and \teff fixed to asteroseismically-determined values. It is necessary to calibrate such a study against benchmark stars with very precisely-determined parameters, which in practice requires nearby bright stars that are amenable to very high signal-to-noise spectroscopy plus asteroseismology \citep{creeveybenchmark}, parallaxes \citep{hawkinsbenchmarks}, and/or interferometry \citep{casagrandebenchmark,creeveybenchmark2}. This is especially important in the context of the \gaia mission \citep{gaia}, which has recently put out its second data release of 1,692,919,135 sources, including 1,331,909,727 with parallaxes \citep{gaiadr2}. These data will form the basis of many large surveys and it is vital that they are calibrated correctly. To this end, 36~FGK stars including both giants and dwarfs have been chosen as \gaia benchmark stars for which metallicities \citep{gaiabenchmark1,2018RNAAS...2c.152J}, effective temperatures and asteroseismic surface gravities \citep{gaiabenchmark3}, and relative abundances of $\alpha$ and iron-peak elements \citep{gaiabenchmark4} have been determined. This includes only four main-sequence stars much cooler than the Sun, due to the paucity of such stars with asteroseismology. This has been accompanied by the release of high-resolution spectra \citep{gaiabenchmark2} and formed the basis of extensions to lower metallicities \citep{gaiabenchmark5}, stellar twin studies \citep{gaiabenchmarktwins} and comparisons of stellar abundance determination pipelines \citep{gaiabenchmarkabundances}. Furthermore, by combining asteroseismology with optical interferometry, it has been possible to determine fundamental parameters of main-sequence and giant stars with unprecedented precision \citep{huber12,thetacygwhite,white15}. 

Brighter \kepler stars are therefore ideal benchmark targets, since photometry can be most easily complemented by \gaia parallaxes, interferometric diameters, and high-resolution spectroscopy.  
Unfortunately, the \kepler field was deliberately placed to minimize overall the number of extremely bright stars on the detectors, so that only a dozen stars brighter than~6th~magnitude landed on silicon \citep{2010ApJ...713L..79K}. This was because stars brighter than $Kp \sim 11$ saturated the CCD detectors, with their flux distributed along a bleed column and rendering those pixels otherwise unusable. Furthermore, due to the limited bandwidth to download data from the spacecraft, only $\sim 5.7\%$ of pixels on the \kepler detectors were actually downloaded in each Quarter \citep{2010ApJ...713L..87J}. The result of these two target selection constraints is that photometry was obtained for most of the mission for only~35 stars brighter than $Kp<7$ in the \kepler field, while a further~17 targets in this range were observed for less than half the mission and~29 targets brighter than this threshold were entirely ignored. The availability of \kepler data remains significantly incomplete down to fainter magnitudes, and in this work we consider $Kp=9$ to be an arbitrary cutoff for bright stars of interest. In the \ktwo mission \citep{k2early}, very saturated stars have been observed with `halo photometry' using unsaturated pixels in a specially-determined region around bright stars, including the Pleiades \citep{halo}, Aldebaran \citep{aldebaran}, $\iota$~Librae \citep{Buysschaert2018}, and $\rho$~Leonis \citep{rholeo}. Unfortunately, in the four-year \kepler sample, photometry of such saturated stars was rarely attempted, with some exceptions, such as RR~Lyrae \citep{Kolenberg2011} and $\theta$~Cyg and 16~Cyg~AB \citep[e.g.][]{thetacygwhite,Guzik2016}.

\citet{orig_smear} noted a way to obtain photometry of every target on-silicon in \kepler using a data channel normally used for calibration, even if active pixels were not allocated and downloaded. Because the \kepler camera lacks a shutter, the detector is exposed to light during the readout process, with the result that fluxes in each pixel are contaminated by light collected from stars in the same column. This is a particularly serious issue for faint stars in the same detector column as brighter stars, and it is important to calibrate this at each readout stage. Twelve rows of blank `masked' pixels were allocated in each column to measure the smear bias; furthermore, twelve `virtual' rows were recorded at the end of the readout, with the result that twelve rows of pixels sample the smear bias in each column. \citet{orig_smear} realized that these encode the light curves of bright targets in a 1D projection of the star field. Compared to the flux in the science image for a given target, the masked and virtual smear rows each receive an incident flux in proportion to the relative exposure times of the smear versus imaging pixels:  $\sim (0.52\,\text{s}/1070\,\text{rows})/6.2\,\text{s}$. If the smear flux is dominated by the light from a single star, the combined flux from the 24 smear rows is equivalent to the normal flux of a star $\sim 6.8$ magnitudes fainter.

In \citet{smear}, we demonstrated a method for extracting precise light curves of bright stars in \kepler and \ktwo from these collateral data, and presented light curves of a small number of variable stars as examples to illustrate this method. In this paper we present smear light curves of all unobserved or significantly under-observed stars brighter than $Kp=9$ in the \kepler field. This sample mostly consists of red giants and hot stars, containing only one~G dwarf. We find no transiting planets, but detect one new eclipsing binary, and measure solar-like oscillations in 33~red giants. We do not model the hot main-sequence stars in great detail, but provide some discussion and initial classification of interesting variability. For the oscillating red giants that constitute the bulk of the sample, we determine the asteroseismic parameters \numax and \Dnu, and therefore stellar masses and \logg measurements. We have also obtained high-resolution optical spectroscopy of 63~stars, predominantly giants, with the Tillinghast Reflector \'{E}chelle Spectrograph \citep[TRES;][]{2007RMxAC..28..129S}. For the~33 stars with both spectroscopy and asteroseismic parameters we derive fundamental stellar parameters and elemental abundances. These asteroseismic constraints can be compared to those from \gaia, offering the opportunity both to test asteroseismic scaling relations and combine both datasets to refine the benchmark star properties further.

We have made all new data products and software discussed in this paper publicly available, and encourage interested readers to use these in their own research.   

\section{Method}
\label{method}

\subsection{Sample}
\label{sample}

We selected all stars on-silicon in \kepler with $Kp<9$ that were targeted for fewer than $8$ quarters. The majority of these were not previously targeted at all, but sixteen stars were to some extent observed conventionally; these are listed in Table~\ref{quarters}. A number of these lay at the edge of a detector, with the result that in some cadences the centroid of the star did not lie on the chip; light curves from these targets were found to be of extremely low quality and all of these stars were discarded. After applying these criteria we obtained a list of 102 targets, which are listed in Table~\ref{all_stars} with their \kepler magnitude $Kp$, together with their spectral type from SIMBAD, \gaia~DR2 apparent~$G$ magnitudes and $Bp-Rp$ colors, \gaia~DR2 calibrated distances from \citet{gaiadists}, variability classification and availability of TRES spectroscopy. It should be noted that $Bp-Rp$ colours are not calibrated for reddening and are therefore not a precise measure of stellar temperature. The \kepler spacecraft rotates between quarters, so that it cycles through four orientation `seasons' each rotated from the last by $90^{\circ}$. Some stars were not on silicon for all seasons: we have only one season of HD~179394; two for HD~187277, HD~226754, V554~Lyr, and BD+47~2891; and three for BD+43~3064. The addition of our sample to the conventionally-observed stars makes the \kepler survey magnitude-complete down to $Kp=9$ for all stars on-silicon.

\begin{deluxetable}{cc}
\tablecaption{Targets observed conventionally for one or more quarters.}
\tablehead{\colhead{Object} & \colhead{Quarters}}
\startdata
HD~174020 & Q2, 6, 10, 14            \\
HD~175841 & Q11-12, 14-16, SC Q3     \\
HD~176582 & Q12-13                   \\
HD~178090 & Q1, 3, 10                \\
HD~180682 & Q0, 3, 7                 \\
HD~181069 & Q1, 10, 13, 14, 17       \\
HD~181878 & Q14-17                   \\
HD~182694 & Q2                       \\
HD~183124 & Even Quarters            \\
HD~185351 & Q1-3; SC Q16             \\
HD~186155 & Q1                       \\
HD~187217 & Q14-17                   \\
HD~188252 & Q13                      \\
HD~189013 & SC Q3                    \\
V380~Cyg  & Q11; SC Q7, 9, 10, 12-17 \\
V819~Cyg  & Q14, 16, 17    \\
\enddata
\tablecomments{Some smear targets were observed conventionally for one or more quarters. SC denotes quarters that were observed in short cadence mode and all others in long cadence.  \label{quarters}}
\end{deluxetable}

\begin{figure*}
\plotone{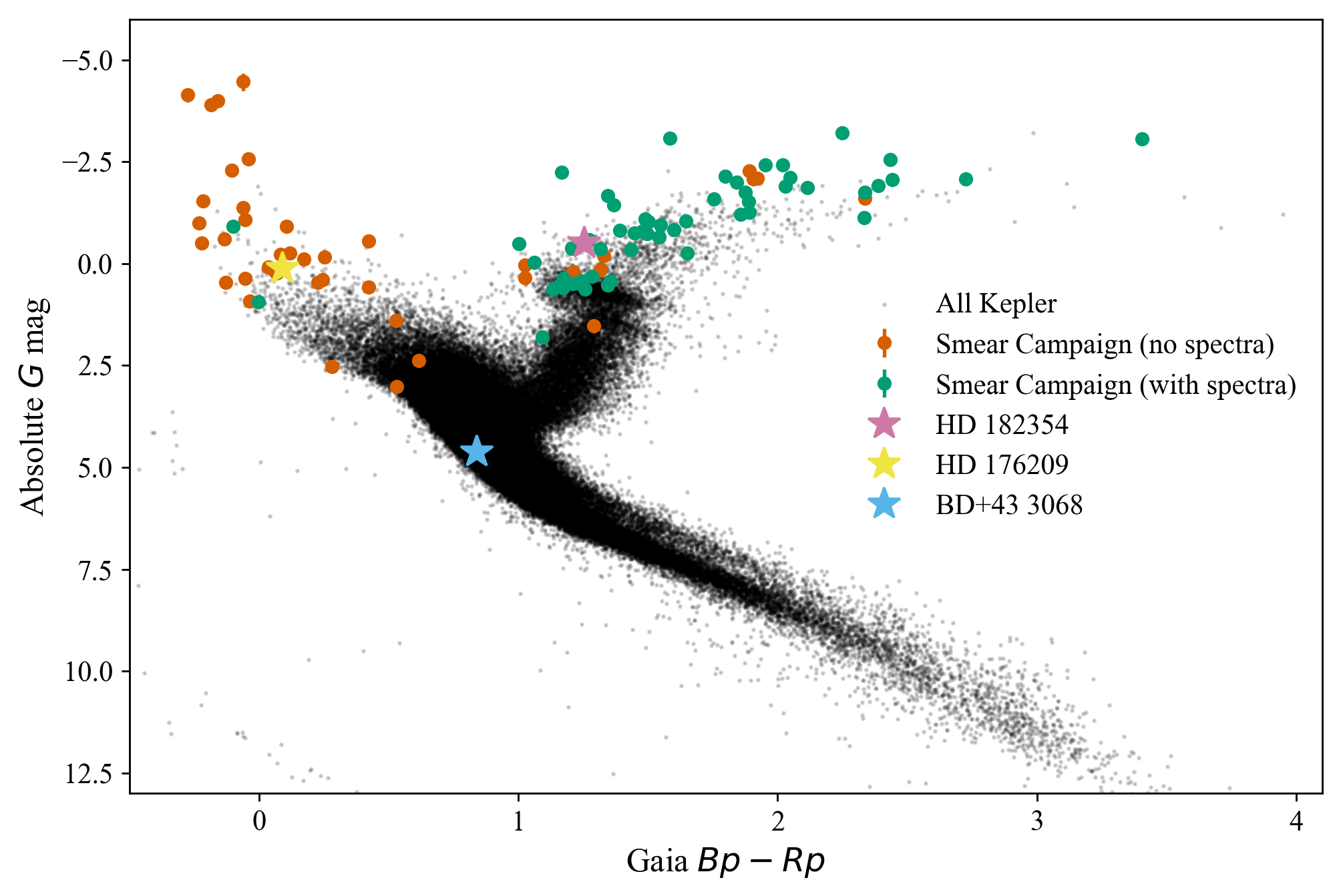}

\caption{\label{hrdiagram}
\gaia colour-magnitude diagram of the Smear Campaign stars (orange and teal) overlaid on the sample of \kepler stars with \gaia parallax $\text{SNR} > 25$ (black), using the Bedell \href{https://gaia-kepler.fun}{\nolinkurl{gaia-kepler.fun}} crossmatch and \gaia~DR2 calibrated distances from \citet{gaiadists}. The smear sample includes giants and hot main-sequence stars. Those giants for which TRES spectroscopy have been obtained are highlighted in teal. Three stars discussed in the text are marked with $\star$ symbols. An interactive version of this diagram is available as supplementary material from the journal or at  \href{https://benjaminpope.github.io/data/cmd_smear.html}{\nolinkurl{benjaminpope.github.io/data/cmd_smear.html}}.}
\end{figure*}

Figure~\ref{hrdiagram} shows these stars on a colour-magnitude diagram using \gaia $Bp-Rp$ and absolute~$G$ magnitudes and \gaia~DR2 calibrated distances \citep{gaiadists}, overlaid on the \kepler sample from the Bedell \href{https://gaia-kepler.fun}{\nolinkurl{gaia-kepler.fun}} crossmatch. The smear targets in this diagram selected to have higher apparent brightnesses than the general \kepler population, appear also to have higher intrinsic luminosities. While this could simply arise from being selected for their apparent brightness, it is {}worth considering whether this is because of a bias in their parallax measurements. While \gaia parallaxes for very bright stars can be subject to systematic error, we have compared to \hipparcos \citep{vanleeuwen07b}, and found close agreement for the brightest stars, with a scatter that increases with magnitude. We therefore suggest that parallax bias is not the reason for the smear sample sitting above the majority of the \kepler sample.

We identify the evolutionary state of main-sequence versus evolved stars from the \gaia colour-magnitude diagram in Figure~\ref{hrdiagram}. Taking a cutoff in \gaia $Bp-Rp > 1$, we identify 66 of these stars as evolved systems, and the remaining~36 lie on the main-sequence. 

The coolest main-sequence star, BD+43~3068 (SAO~47785), is a G0 dwarf with a $G$ magnitude of 8.3 and a distance of $53.8 \pm 0.1$~pc, and it is therefore surprising that it was not included in the nominal \kepler survey as a solar analogue. It is only possible to reconstruct a light curve with the 30~minute long cadence and therefore it is not possible to do asteroseismology on this bright, nearby solar-like star. Its light curve shows neither rotational modulation (as determined by its featureless autocorrelation) nor evidence for transits.

Considering stars lying close to the main-sequence, from the \kepler power spectrum we identify solar-like oscillations in HD~182354 and HD~176209 at frequencies consistent with them being subgiants or contaminated with flux from red giants.

\subsection{Photometry}
\label{photometry}

In generating light curves of the \kepler smear stars, we have followed the methods described by \citet{smear}, with some improvements. We selected our input RA and Dec values from the \kepler Input Catalog (KIC) \citep{kic}, and queried MAST to find the corresponding mean pixel position for each \kepler quarter. We then measured the centroid of smear columns in the vicinity, and used these values to do raw aperture photometry. We found that the cosine-bell aperture used for raw photometry by \citet{smear} can sometimes introduce position-dependent systematics and jumps. We instead used a super-Gaussian aperture, 

\begin{equation}
A \propto \exp{\dfrac{-(x-x_0)}{w} ^ 4},
\end{equation}
\noindent where $x_0$ is the centroid and $w$ is the width in pixels. The very flat top of this function helps avoid significant variation with position, while still smoothly rolling off at the edges to avoid discontinuous artefacts. The super-Gaussian is calculated on a grid of $10\,\times$ subsampled points in pixel space so that the sharply varying edge changes column weights smoothly as a function of centroid. We then extracted photometry using apertures with a range of widths $w \in\{1.5,~2,~3,~4,~5\}$ pixels.

From this raw photometry a background light curve was subtracted, correcting for time-varying global systematics. Whereas in \citet{smear} we subtracted a background estimated manually, for this larger set of light curves, we have chosen the lowest $25\%$ of pixels by median flux as being unlikely to be contaminated by stars, and taken our background level to be the median of this at each time sample. To reduce the noise in this background model, we fitted a Gaussian Process (GP) with a 30-day timescale squared exponential kernel using \textsc{george} \citep{hodlr}, and our final background light curve is taken to be the posterior mean of this GP. 

The dominant source of residual systematic errors in nominal \kepler time series is a common-mode variation primarily due to thermal changes on board the spacecraft, an issue that is traditionally dealt with by identifying and fitting a linear combination of systematic modes \citep{pdc0,pdc1,pdc2,petigura}. We have adopted the same approach here, using the \kepler Pre-search Data Conditioning (PDC) Cotrending Basis Vectors (CBVs) available from MAST, finding least-squares fits of either the first~4 or~8 CBVs to each light curve. This can remove astrophysical signals having long timescales, and so we use and recommend 4~CBV light curves for stars with variability on timescales longer than $\sim 5$ days, or indeed raw uncorrected light curves for stars variable at high amplitude on $\sim$~quarter timescales, but otherwise we recommend the 8~CBV light curves. There is some room for improvement here by simultaneously modelling astrophysical and instrumental variations, but this is beyond the scope of this paper. In the following, we use the light curves with the lowest 6.5~hr Combined Differential Photometric Precision (CDPP) \citep{cdpp} out of all apertures, as calculated with the \textsc{k2sc} CDPP implementation \citep{k2sc}. This is not necessarily the optimal choice for all red giants, especially those with oscillations on a 6.5~h timescale, but is a reasonable proxy for white noise and leads to satisfactory results upon visual inspection of the present sample.

\begin{figure}
\plotone{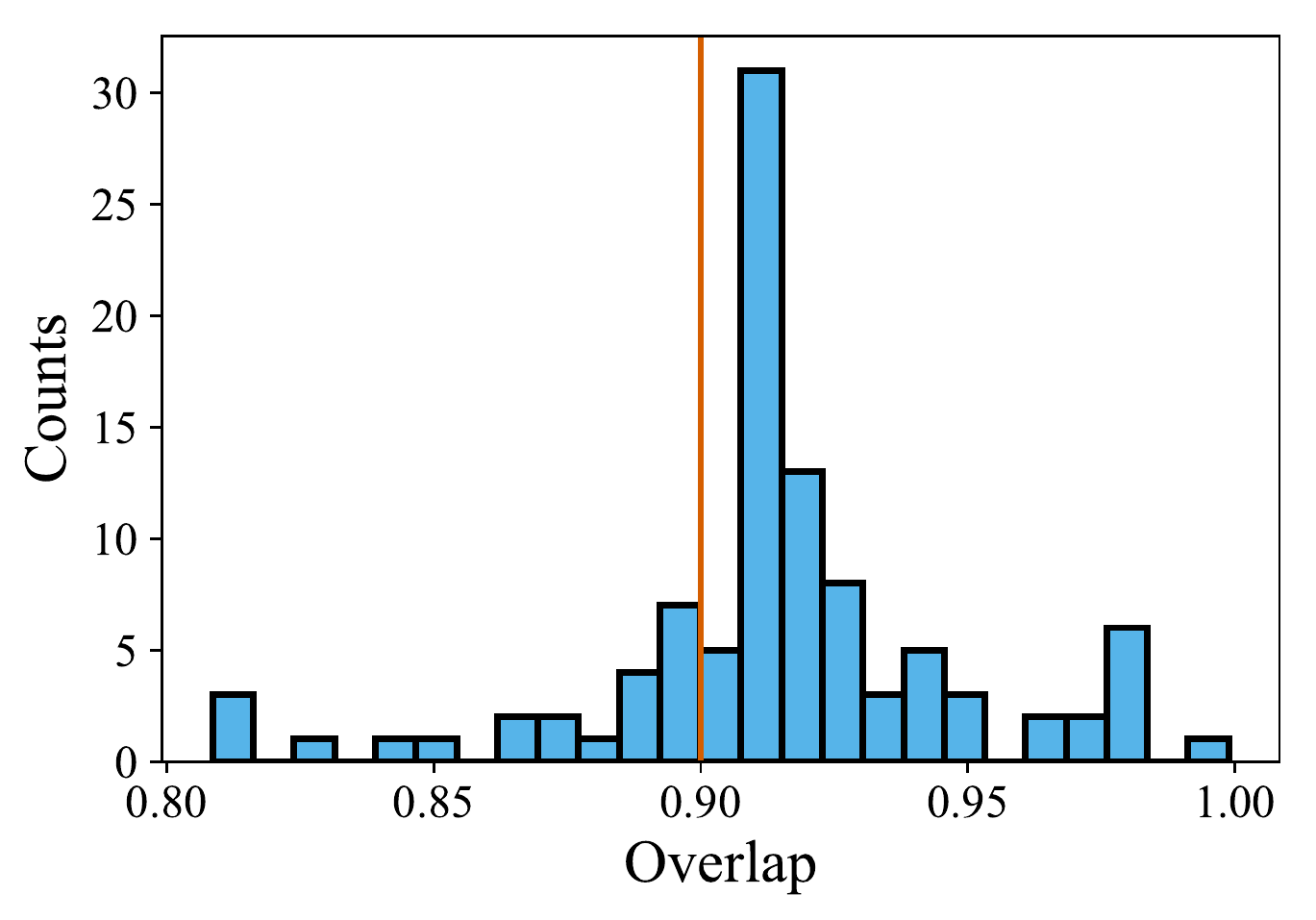}
\caption{\label{contamination2}
Histogram over overlap integrals of smoothed periodograms of odd and even quarters for each star in the sample. The peak at $\sim 0.91$ contains normal stars with limited contamination; we investigate the 22 stars with overlaps below 0.9 for which there is a significant possibility of contamination.}
\end{figure}

We can assess the importance of this contamination by considering differences between quarters. Because the Kepler spacecraft rotates $90^\circ$ between successive quarters, stars were observed on different CCD modules with the exception of stars on the central Module~13. Minor variations in the precise alignment of each CCD mean that the contribution from contaminating stars varies from quarter to quarter. Differences are clearer for Module~13, where contaminating stars will only be aligned along the same columns as a smear target every second quarter. We have therefore generated Lomb-Scargle periodograms \citep{lomb,scargle} of each light curve, after clipping for outliers. We consider only odd and even quarters separately, and also the full combined time series. In the great majority of cases they closely resemble one another, indicating that contamination is at worst a minor effect. In order to better quantify this, we computed the inner product of normalized periodograms of the odd and even quarters, each smoothed with a 3-element Gaussian kernel. If this overlap integral is 1, then the power spectra are identical; substantial departures from unity may be caused by real nonstationary or long-period stellar variation, noise, or gain or contamination differences between the seasons. We found that the distribution of overlaps (Figure~\ref{contamination2}) is strongly peaked around $\sim 0.91$, with a tail of 22 stars showing overlap $< 0.9$. We investigate these further, finding that in some of these cases  there was no obvious problem. For example, the classical pulsator HD~175841 showed amplitude changes of several pulsations but the overall distribution seemed very similar, from which we conclude that the variation is probably astrophysical \citep[as in][]{2016MNRAS.460.1970B}. 

\begin{figure}
\plotone{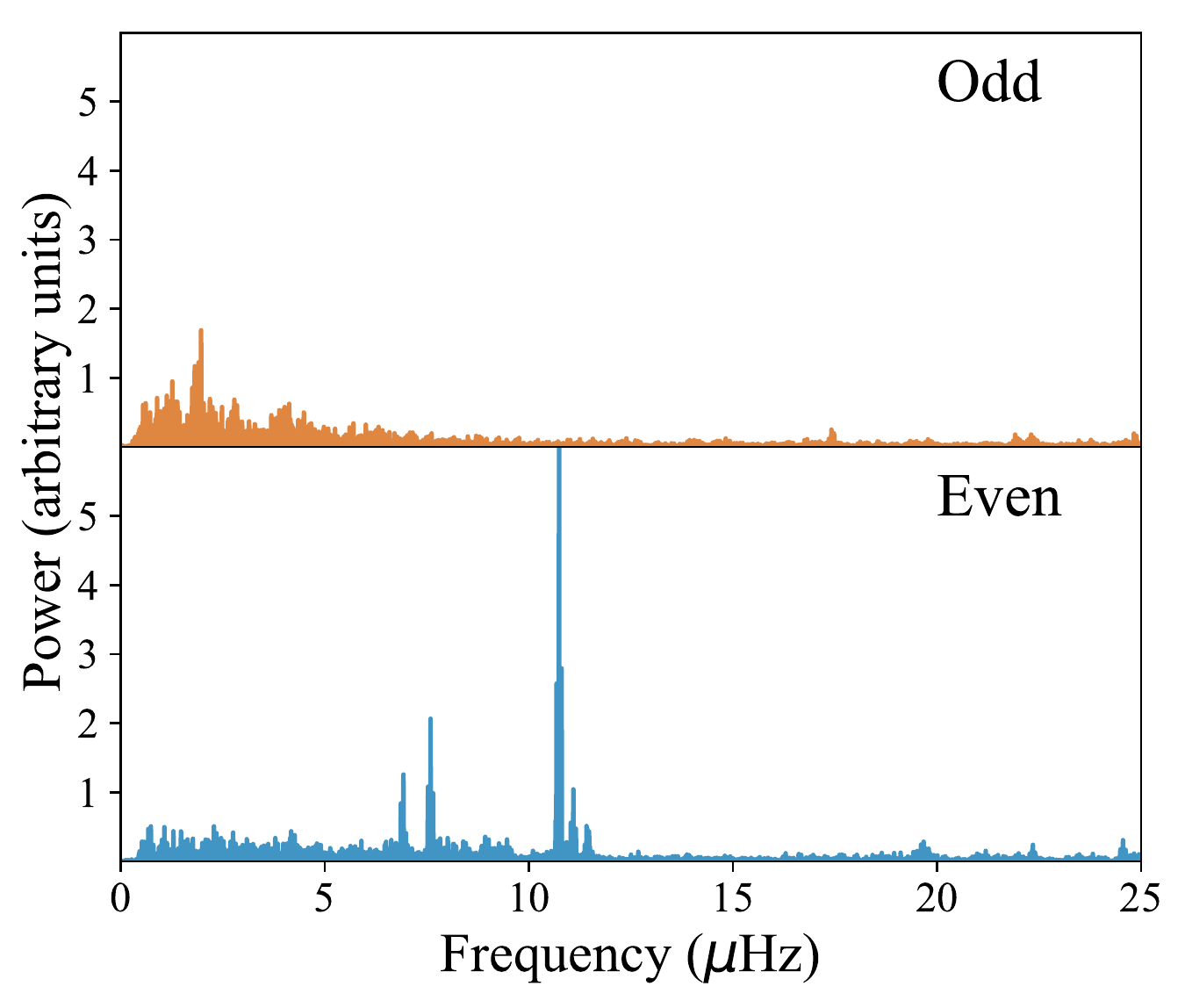}
\caption{\label{contamination}
Power spectra of odd and even quarters of HD~181778. Even quarters have very high amplitude coherent oscillations that are absent in odd quarters.}
\end{figure}

\begin{figure}
\plotone{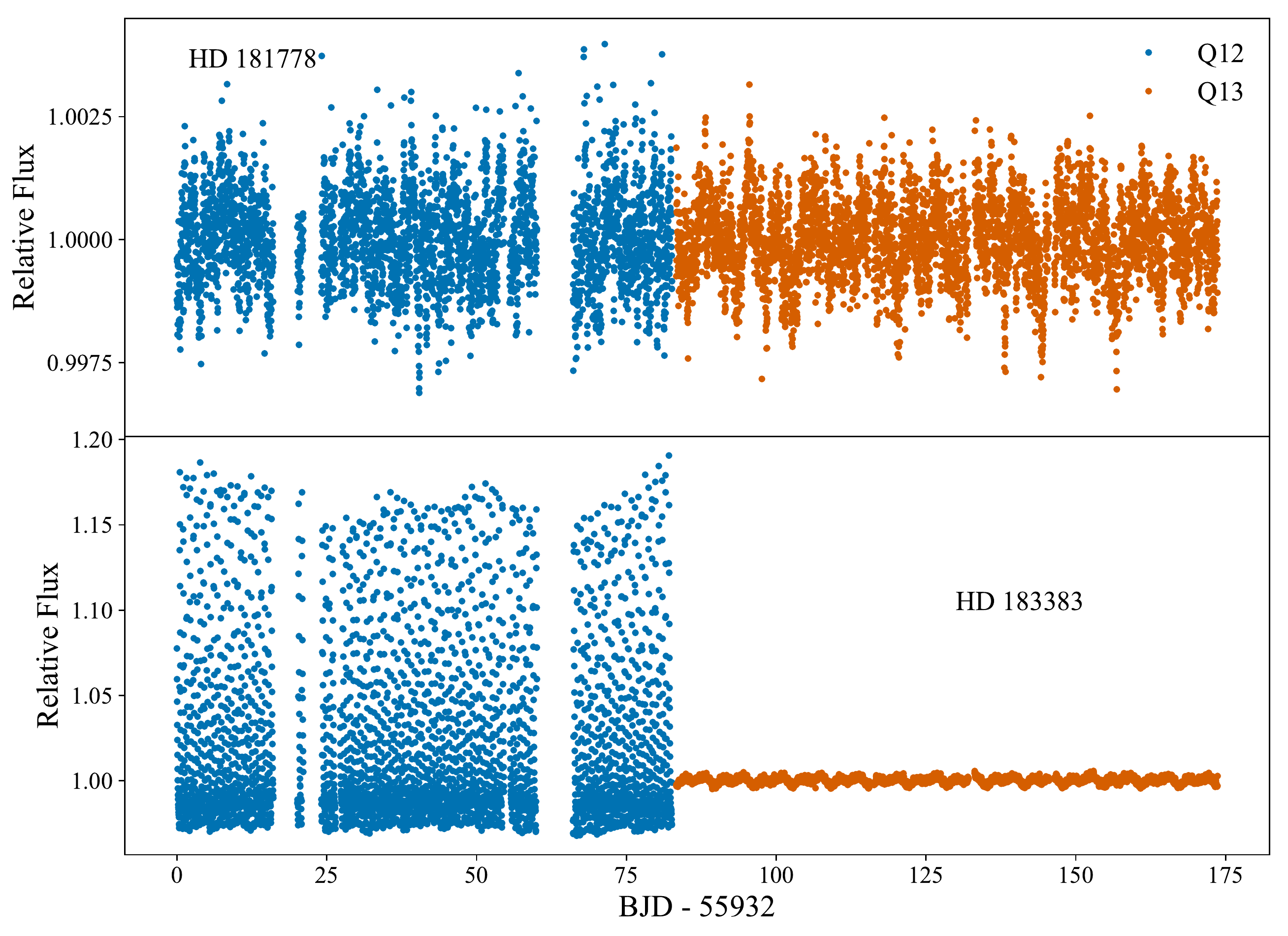}
\caption{\label{contamination_timeseries}
Time series of Quarters 12-13 of HD~181778 and HD~183383, both of which show contamination. For HD~181778, contamination is not apparent to the eye in the time series, but Figure~\ref{contamination} showed that in its power spectrum there is a significant effect from a coherent oscillator. Meanwhile in HD~183383, even quarters show easily visible contamination from the star RR~Lyr, some quarters worse than others, while odd quarters show low amplitude coherent variability consistent with an ellipsoidal variable.} 
\end{figure}

In the case of HD~181878, a red giant on Module~13, there is clear and significant contamination from a star with several low-frequency pulsations, as is seen in Figure~\ref{contamination}. Likewise the light curve for HD 183383, which variously falls on Modules~8, 12, 14 and~18, shows different behaviour for different quarters: some parts are likely from an ellipsoidal variable with a period of 6.46 days, while other quarters are contaminated by the star RR~Lyrae. The effects of this contamination in the time domain are shown in Figure~\ref{contamination_timeseries}: there is very little discernable effect for HD~181878 by eye, whereas the RR~Lyr contamination in HD~183383 is readily apparent. Between seasons, there is an extra hump of power near the red giant oscillations in HD~175740; extra low frequency power in HD~180658; one coherent peak in  HD~182694; high frequency contamination in HD~181597 possibly from an EB; and a very significant difference in amplitude between seasons for BD~+39~3882. In other cases visual inspection does not show severe contamination, but in all cases we recommend that users of these light curves carefully check for differences between quarters, as well as investigating the full frame images for potential contaminants.

\subsection{Asteroseismology}
\label{asteroseismology}

Among the~66 red giants identified in this sample, for~22 the timescale of their variability is similar to the length of a \kepler quarter and they are thus badly affected by systematics which are hard to correct with the CBV approach. In Table~\ref{all_stars} we have noted these as `long-period variables' (LPVs). For the~33 giants with high-SNR shorter-timescale variability, we have attempted to extract the asteroseismic parameters \numax and $\langle \Dnu \rangle$ \citep{KB95,2013ARA&A..51..353C}. These constrain fundamental stellar parameters through the approximate scaling relations: 

\begin{equation}
\label{scaling}
\numax \propto \dfrac{g}{g_{\sun}} \cdot \left(\dfrac{\teff}{\teff_{\sun}}\right)^{\dfrac{1}{2}},
\end{equation}

and

\begin{equation}
\label{Dnuscaling}
\langle \Dnu \rangle \propto \sqrt{\langle \rho \rangle} = \sqrt{M {R}^{-3}}.
\end{equation}

We have followed the method of \citet{2016AN....337..774D}, obtaining a Lomb-Scargle periodogram of the smoothed time series according to the method of \citet{2011MNRAS.414L...6G}. The posterior distribution of the asteroseismic parameters was obtained with a Markov Chain Monte Carlo fit to the smoothed periodogram, applying the combined granulation and oscillation model of \citet{2014A&A...570A..41K}. This consists of two Harvey profiles for the granulation \citep{1985ESASP.235..199H}, a Gaussian envelope for the stellar oscillations, and a white noise background for photon noise. The marginal posterior distribution for the oscillation envelope is well-approximated by a single Gaussian, and we have taken its median and standard deviation to be our estimates for \numax and its uncertainty.

To estimate \Dnu, we divided the power spectrum by the granulation and noise models to obtain a signal-to-noise spectrum, and fit a sum of Lorentzians separated by mean large (\Dnu) and small (\dnu) separations to the part of this spectrum in the vicinity of \numax. For this dataset, \dnu is not constrained, but mean $\langle \Dnu \rangle$ is typically well-constrained and its posterior marginal distribution is well-represented by a single Gaussian. We also fit the dimensionless parameter $\epsilon$, which is the offset of the lowest frequency in the comb of p-modes from zero in units of $\Dnu$.

We obtained good estimates of these asteroseismic parameters for 33 targets, presented in Table~\ref{astero_table}. In the remainder of cases, as noted above, very-low-frequency ($\lesssim 2\muHz$) oscillations are affected by filter artefacts from detrending, and we were not able to obtain good estimates for these stars. 

For eight stars, we found that the asteroseismic fit is unsatisfactory: for BD+39~388 we cannot detect the expected oscillations; for BD+43~3064 there are significant peaks but these are not consistent with the pattern expected from a red giant; for HD~179959 and HD~187217 we suspect contamination with the oscillations of a second giant, which is hard to remove from smear light curves; for HD~188629, HD~188639 and HD~188875 we can extract a \numax but not a robust \Dnu. The `retired~A star' HD~185351 \citep[studied by][]{2014ApJ...794...15J}, has a mode envelope that is not well fit by our model. The smear light curve for this star has already been published by \citet{2017MNRAS.464.3713H}, who showed with detailed asteroseismic modelling that it had a zero-age main-sequence mass of $\sim 1.60 \msun$ (a so-called `retired~A star') and used it to calibrate the convective overshoot parameter for low-luminosity red giants. The global asteroseismic modelling presented here should therefore be considered to be superseded by the more detailed model of \citet{2017MNRAS.464.3713H}. 

\begin{deluxetable}{cccc}
\tablecaption{Global asteroseismic parameters \Dnu, \numax, and $\epsilon$ for the red giant sample as discussed in Section~\ref{asteroseismology}.\label{astero_table}}
\tablehead{\colhead{Object} & \colhead{\Dnu} & \colhead{\numax} & \colhead{$\epsilon$}}
\startdata
BD+36 3564 & $0.95 \pm 0.03$ & $5.08 \pm 0.10$ & $0.83 \pm 0.20$ \\
BD+39 3577 & $1.68 \pm 0.01$ & $13.27 \pm 0.32$ & $0.74 \pm 0.06$ \\
BD+42 3150 & $4.22 \pm 0.03$ & $38.32 \pm 0.96$ & $0.70 \pm 0.07$ \\
BD+43 3171 & $0.42 \pm 0.05$ & $1.98 \pm 0.05$ & $0.80 \pm 0.17$ \\
BD+43 3213 & $0.49 \pm 0.01$ & $2.56 \pm 0.06$ & $1.01 \pm 0.07$ \\
BD+48 2904 & $2.85 \pm 0.01$ & $23.13 \pm 0.72$ & $0.86 \pm 0.08$ \\
BD+48 2955 & $0.90 \pm 0.01$ & $5.44 \pm 0.08$ & $0.81 \pm 0.05$ \\
HD 174020 & $0.56 \pm 0.02$ & $2.48 \pm 0.10$ & $0.89 \pm 0.08$ \\
HD 174829 & $1.28 \pm 0.01$ & $7.95 \pm 0.16$ & $0.78 \pm 0.06$ \\
HD 175740 & $5.93 \pm 0.01$ & $64.33 \pm 0.78$ & $1.00 \pm 0.02$ \\
HD 175884 & $1.12 \pm 0.01$ & $7.07 \pm 0.11$ & $0.96 \pm 0.08$ \\
HD 178797 & $1.03 \pm 0.02$ & $6.34 \pm 0.09$ & $0.74 \pm 0.29$ \\
HD 178910 & $3.64 \pm 0.02$ & $32.06 \pm 0.31$ & $0.83 \pm 0.05$ \\
HD 179396 & $3.76 \pm 0.02$ & $31.02 \pm 0.44$ & $0.92 \pm 0.03$ \\
HD 180312 & $4.17 \pm 0.02$ & $33.84 \pm 0.28$ & $0.96 \pm 0.04$ \\
HD 180475 & $0.82 \pm 0.00$ & $4.34 \pm 0.10$ & $0.68 \pm 0.03$ \\
HD 180658 & $4.00 \pm 0.02$ & $33.76 \pm 0.50$ & $0.90 \pm 0.05$ \\
HD 180682 & $0.77 \pm 0.05$ & $3.68 \pm 0.08$ & $1.07 \pm 0.15$ \\
HD 181022 & $0.38 \pm 0.01$ & $1.58 \pm 0.03$ & $0.70 \pm 0.10$ \\
HD 181069 & $4.43 \pm 0.01$ & $41.46 \pm 0.32$ & $0.90 \pm 0.02$ \\
HD 181097 & $1.61 \pm 0.02$ & $11.16 \pm 0.14$ & $0.72 \pm 0.36$ \\
HD 181597 & $3.11 \pm 0.01$ & $25.84 \pm 0.25$ & $0.97 \pm 0.02$ \\
HD 181778 & $2.56 \pm 0.02$ & $22.86 \pm 0.29$ & $0.72 \pm 0.06$ \\
HD 181880 & $1.04 \pm 0.01$ & $6.54 \pm 0.10$ & $0.76 \pm 0.05$ \\
HD 182354 & $2.66 \pm 0.01$ & $24.73 \pm 0.37$ & $0.74 \pm 0.04$ \\
HD 182531 & $1.03 \pm 0.00$ & $6.47 \pm 0.09$ & $0.86 \pm 0.03$ \\
HD 182692 & $4.66 \pm 0.01$ & $44.38 \pm 0.47$ & $0.87 \pm 0.02$ \\
HD 182694 & $5.71 \pm 0.01$ & $69.78 \pm 1.02$ & $0.94 \pm 0.25$ \\
HD 183124 & $4.39 \pm 0.01$ & $39.59 \pm 0.29$ & $0.95 \pm 0.03$ \\
HD 185286 & $0.72 \pm 0.01$ & $4.23 \pm 0.10$ & $0.73 \pm 0.08$ \\
HD 188537 & $1.55 \pm 0.01$ & $13.40 \pm 0.34$ & $0.72 \pm 0.07$ \\
HD 189750 & $4.16 \pm 0.04$ & $36.14 \pm 0.58$ & $0.94 \pm 0.08$ \\
HD 226754 & $1.19 \pm 0.01$ & $7.41 \pm 0.19$ & $0.74 \pm 0.08$
\enddata
\end{deluxetable}

\subsection{Spectroscopy}
\label{spectroscopy}

We have obtained high-resolution spectroscopy with TRES for~63 stars, mainly giants, in order to constrain stellar parameters and elemental abundances. Operating with spectral resolving power $R=44 000$, we have obtained spectra with a mean signal-to-noise ratio (SNRs) of $\sim 100$  per resolution element. 
From this observing run we have~33 unique targets with seismic \logg and spectra, a number comparable to the 36 of the \gaia benchmark set \citep{2018RNAAS...2c.152J} and a significant addition to the ensemble of bright red giants with asteroseismic parameter determinations. 

We have used Equation~\ref{scaling}, the asteroseismic scaling relation for \numax \citep{1991ApJ...368..599B,KB95}, to estimate \logg in order to inform extraction of chemical abundances from spectra. Using the initial spectroscopic estimate of \teff, which is not dependent on \numax, uncertainties in \numax were propagated with Monte Carlo sampling. 

To derive stellar parameters from our TRES spectra, we initially ran the Stellar Parameter Classification code \citep[SPC:][]{spc} to determine \teff and \logg, using the SPC \teff to inform the asteroseismic estimation of \logg from \numax. For deriving abundances, \teff was fixed from the results of an initial SPC fit, while \logg was fixed to the seismic values. For four stars with low \logg and metallicity (BD+43~3171, HD~174020, HD~180682, and HD~181022), the stellar spectral templates in SPC gave unsatisfactory fits. In these cases, \teff was fixed to the results of a broadband spectral energy distribution (SED) fit to archival photometry as catalogued by \citet{Mcdonald2017}, and \logg calculated from these without iteration.

The other stellar atmospheric parameters including the microturbulent velocity ($v_{\text{mic}}$), broadening (convolution by $v_{\text{mac}}$, $v_{\sin{i}}$ and the instrumental line profile), as well as [Fe/H] and chemical abundances for 13 chemical species were derived using the Brussels Automatic Code for Characterizing High accUracy Spectra \citep[BACCHUS:][]{bacchus}. The results from this calculation are displayed in Table~\ref{stellar_props}. BACCHUS uses an interpolation scheme through a grid of MARCS model atmospheres \citep{Gustafsson2008} in combination with TURBOSPECTRUM \citep{Alvarez1998, Plez2012}. For the calculation of synthetic spectra, atomic line information has been taken from the fifth version of the Gaia-ESO linelist (Heiter et al., in preparation). Additionally, we used the molecular species for CH \citep{Masseron2014}, CN, NH, OH, MgH  C$_{2}$ (T. Masseron, private communication). The SiH molecular information was adopted from the Kurucz linelists and the information for TiO, ZrO, FeH, CaH from B. Plez (private communication). 

Individual elemental abundances were derived by first fixing the stellar atmospheric parameters to those determined above. Spectra were then synthesized in regions centered around an absorption feature of the element in question with different [X/Fe] values. A $\chi^2$ minimization procedure was then done to derive the best fitting abundance for each line. The reported abundances are the median [X/Fe] value of the various line regions for each element. Abundance uncertainties reported are the standard error in the line-by-line abundance ratios. Where only one line exists for a given element, we assumed the standard error to be 0.10~dex. In principle, these uncertainties are underestimated because there they do not include the errors driven by imperfect stellar parameters and other systematic errors arising, for instance, from incorrect line list data. We do note, however, the use of asteroseismology to determine \logg greatly reduces the uncertainties caused by the stellar parameters \citep[see][for a longer discussion on this]{hawkinsapogee}. 

To achieve the most precise abundances we have derived them both with and without a line-by-line differential approach with respect to Arcturus ($\alpha$~Bo\"{o}tis), using the method described by \citet{hawkinsapogee} and the Arcturus abundances from \citet{gaiabenchmark4}. Choosing this method means we do not derive the abundances for neutron capture  elements (e.g. Sr, Y, Zr, Ba, La, Nd, Eu) in a differential way because there are no estimated values for these elements at the appropriate benchmark parameters of Arcturus. For these elements we instead derived the chemical abundances in an absolute way, where the solar abundances of \cite{Asplund2005} were assumed. The uncertainty in the abundances are either the line-by-line dispersion, or assumed to be 0.10 when just one line is available. No abundances for oxygen could be reliably derived for any of the stars in our spectroscopic sample by either method.

\begin{deluxetable}{ccccccccc}
\tablecaption{Fundamental stellar parameters for the red giant sample as determined jointly by asteroseismology (asteroseismic \logg; Section~\ref{asteroseismology}) and spectroscopy (RV, \teff, \logg, [M/H], $V\sin{i}$, Mass, Radius, and Age; Section~\ref{spectroscopy}.)\label{stellar_props}}
\tablehead{\colhead{Object} & \colhead{RV} & \colhead{\teff} & \colhead{\logg} & \colhead{[M/H]} & \colhead{$V\sin{i}$} & \colhead{Mass} & \colhead{Radius} & \colhead{Age}}
\startdata
BD+36 3564 & $-77.84 \pm 0.05$ & $4100 \pm 50$ & $1.5696 \pm 0.0085$ & $-0.63 \pm 0.08$ & $5.54 \pm 0.50$ & $0.91^{+0.10}_{-0.06}$ & $25.61^{+1.25}_{-0.83}$ & $12.40^{+3.60}_{-3.90}$ \\
BD+39 3577 & $-14.81 \pm 0.07$ & $4737 \pm 50$ & $2.0178 \pm 0.0103$ & $-0.41 \pm 0.08$ & $4.78 \pm 0.50$ & $2.39^{+0.22}_{-0.19}$ & $24.78^{+0.88}_{-0.72}$ & $0.65^{+0.20}_{-0.19}$ \\
BD+42 3150 & $-26.52 \pm 0.07$ & $4776 \pm 50$ & $2.4804 \pm 0.0108$ & $-0.19 \pm 0.08$ & $4.22 \pm 0.50$ & $1.42^{+0.14}_{-0.14}$ & $11.27^{+0.39}_{-0.41}$ & $2.90^{+1.30}_{-0.70}$ \\
BD+43 3171 & $-16.32 \pm 0.11$ & $3656 \pm 50$ & $1.1365 \pm 0.0112$ & $-1.20 \pm 0.08$ & $4.54 \pm 0.50$ & $1.07^{+0.31}_{-0.17}$ & $45.24^{+6.08}_{-3.73}$ & $7.90^{+7.00}_{-4.60}$ \\
BD+43 3213 & $-14.16 \pm 0.16$ & $3901 \pm 50$ & $1.2619 \pm 0.0106$ & $-0.16 \pm 0.08$ & $6.82 \pm 0.50$ & $1.59^{+0.14}_{-0.14}$ & $48.51^{+1.92}_{-1.87}$ & $2.40^{+0.80}_{-0.60}$ \\
BD+48 2904 & $5.24 \pm 0.03$ & $4484 \pm 50$ & $2.2474 \pm 0.0137$ & $-0.30 \pm 0.08$ & $4.11 \pm 0.50$ & $1.28^{+0.13}_{-0.12}$ & $14.13^{+0.45}_{-0.45}$ & $4.40^{+1.70}_{-1.20}$ \\
BD+48 2955 & $1.66 \pm 0.04$ & $4143 \pm 50$ & $1.6018 \pm 0.0066$ & $-0.60 \pm 0.08$ & $5.33 \pm 0.50$ & $1.60^{+0.10}_{-0.08}$ & $32.71^{+0.82}_{-0.86}$ & $1.80^{+0.30}_{-0.30}$ \\
HD 174020 & $-14.84 \pm 0.08$ & $3781 \pm 50$ & $1.2677 \pm 0.0170$ & $-1.03 \pm 0.08$ & $5.38 \pm 0.50$ & $0.98^{+0.14}_{-0.08}$ & $38.44^{+2.42}_{-1.63}$ & $12.40^{+4.90}_{-4.80}$ \\
HD 174829 & $10.15 \pm 0.03$ & $4381 \pm 50$ & $1.7789 \pm 0.0087$ & $-0.48 \pm 0.08$ & $4.71 \pm 0.50$ & $1.32^{+0.10}_{-0.09}$ & $24.35^{+0.66}_{-0.62}$ & $3.30^{+0.90}_{-0.60}$ \\
HD 175740 & $-8.81 \pm 0.04$ & $4875 \pm 50$ & $2.7099 \pm 0.0053$ & $-0.12 \pm 0.08$ & $3.90 \pm 0.50$ & $1.78^{+0.02}_{-0.01}$ & $9.70^{+0.03}_{-0.04}$ & $1.60^{+0.20}_{-0.00}$ \\
HD 175884 & $-34.39 \pm 0.07$ & $4306 \pm 50$ & $1.7240 \pm 0.0070$ & $-0.41 \pm 0.08$ & $4.91 \pm 0.50$ & $1.57^{+0.09}_{-0.09}$ & $28.14^{+0.66}_{-0.69}$ & $2.00^{+0.50}_{-0.30}$ \\
HD 178797 & $6.35 \pm 0.05$ & $4201 \pm 50$ & $1.6711 \pm 0.0065$ & $-0.63 \pm 0.08$ & $4.82 \pm 0.50$ & $1.44^{+0.13}_{-0.13}$ & $28.43^{+1.16}_{-1.06}$ & $2.50^{+0.90}_{-0.60}$ \\
HD 178910 & $-14.28 \pm 0.05$ & $4560 \pm 50$ & $2.3930 \pm 0.0041$ & $0.12 \pm 0.08$ & $4.38 \pm 0.50$ & $1.45^{+0.05}_{-0.06}$ & $12.53^{+0.17}_{-0.22}$ & $3.40^{+0.60}_{-0.50}$ \\
HD 179396 & $24.80 \pm 0.04$ & $4731 \pm 50$ & $2.3867 \pm 0.0062$ & $-0.24 \pm 0.08$ & $4.32 \pm 0.50$ & $1.21^{+0.05}_{-0.06}$ & $11.52^{+0.19}_{-0.20}$ & $4.90^{+0.80}_{-0.70}$ \\
HD 180312 & $-21.94 \pm 0.05$ & $4868 \pm 50$ & $2.4307 \pm 0.0035$ & $-0.49 \pm 0.08$ & $4.25 \pm 0.50$ & $1.07^{+0.04}_{-0.03}$ & $10.33^{+0.16}_{-0.13}$ & $6.30^{+1.30}_{-0.80}$ \\
HD 180475 & $-45.90 \pm 0.08$ & $4129 \pm 50$ & $1.5025 \pm 0.0095$ & $-0.85 \pm 0.08$ & $5.34 \pm 0.50$ & $1.11^{+0.10}_{-0.09}$ & $30.68^{+1.06}_{-1.01}$ & $5.40^{+1.90}_{-1.50}$ \\
HD 180658 & $2.97 \pm 0.06$ & $4717 \pm 50$ & $2.4228 \pm 0.0064$ & $-0.17 \pm 0.08$ & $3.99 \pm 0.50$ & $1.20^{+0.07}_{-0.07}$ & $11.03^{+0.22}_{-0.21}$ & $5.20^{+1.20}_{-0.80}$ \\
HD 180682 & $30.99 \pm 0.07$ & $4077 \pm 50$ & $1.4700 \pm 0.0099$ & $-1.03 \pm 0.08$ & $5.75 \pm 0.50$ & $0.95^{+0.20}_{-0.11}$ & $30.70^{+3.06}_{-1.82}$ & $10.00^{+5.70}_{-5.00}$ \\
HD 181022 & $-80.39 \pm 0.16$ & $3557 \pm 50$ & $1.0487 \pm 0.0084$ & $-1.63 \pm 0.08$ & $4.68 \pm 0.50$ & $1.02^{+0.12}_{-0.10}$ & $49.79^{+2.82}_{-2.49}$ & $8.50^{+4.00}_{-2.90}$ \\
HD 181069 & $9.99 \pm 0.05$ & $4740 \pm 50$ & $2.5131 \pm 0.0033$ & $-0.09 \pm 0.08$ & $3.95 \pm 0.50$ & $1.50^{+0.04}_{-0.03}$ & $11.13^{+0.10}_{-0.09}$ & $2.70^{+0.30}_{-0.30}$ \\
HD 181097 & $-5.60 \pm 0.08$ & $4389 \pm 50$ & $1.9263 \pm 0.0056$ & $-0.39 \pm 0.08$ & $4.50 \pm 0.50$ & $1.48^{+0.10}_{-0.09}$ & $21.61^{+0.60}_{-0.59}$ & $2.50^{+0.60}_{-0.50}$ \\
HD 181597 & $-13.06 \pm 0.04$ & $4612 \pm 50$ & $2.3018 \pm 0.0042$ & $-0.35 \pm 0.08$ & $3.51 \pm 0.50$ & $1.46^{+0.06}_{-0.04}$ & $13.95^{+0.18}_{-0.16}$ & $2.60^{+0.20}_{-0.30}$ \\
HD 181880 & $0.56 \pm 0.08$ & $4200 \pm 50$ & $1.6850 \pm 0.0065$ & $-0.56 \pm 0.08$ & $4.91 \pm 0.50$ & $1.60^{+0.10}_{-0.09}$ & $29.72^{+0.72}_{-0.71}$ & $1.80^{+0.40}_{-0.30}$ \\
HD 182354 & $-36.79 \pm 0.06$ & $4697 \pm 50$ & $2.2867 \pm 0.0065$ & $-0.30 \pm 0.08$ & $5.38 \pm 0.50$ & $2.37^{+0.10}_{-0.14}$ & $18.20^{+0.17}_{-0.42}$ & $0.70^{+0.05}_{-0.10}$ \\
HD 182531 & $-7.34 \pm 0.05$ & $4204 \pm 50$ & $1.6800 \pm 0.0060$ & $-0.49 \pm 0.08$ & $4.94 \pm 0.50$ & $1.63^{+0.10}_{-0.09}$ & $30.08^{+0.73}_{-0.69}$ & $1.80^{+0.40}_{-0.20}$ \\
HD 182692 & $-8.01 \pm 0.04$ & $4762 \pm 50$ & $2.5438 \pm 0.0045$ & $0.03 \pm 0.08$ & $4.55 \pm 0.50$ & $1.48^{+0.04}_{-0.04}$ & $10.70^{+0.10}_{-0.11}$ & $3.20^{+0.30}_{-0.30}$ \\
HD 182694 & $-0.87 \pm 0.06$ & $5089 \pm 50$ & $2.7546 \pm 0.0063$ & $-0.19 \pm 0.08$ & $5.30 \pm 0.50$ & $2.70^{+0.02}_{-0.06}$ & $11.41^{+0.04}_{-0.08}$ & $0.50^{+0.05}_{-0.02}$ \\
HD 183124 & $14.96 \pm 0.02$ & $4781 \pm 50$ & $2.4949 \pm 0.0031$ & $-0.27 \pm 0.08$ & $5.51 \pm 0.50$ & $1.38^{+0.03}_{-0.05}$ & $10.89^{+0.07}_{-0.16}$ & $3.10^{+0.50}_{-0.30}$ \\
HD 185286 & $-13.70 \pm 0.08$ & $4090 \pm 50$ & $1.4894 \pm 0.0104$ & $-0.37 \pm 0.08$ & $5.98 \pm 0.50$ & $1.66^{+0.08}_{-0.13}$ & $38.30^{+0.80}_{-1.18}$ & $1.90^{+0.50}_{-0.30}$ \\
HD 188537 & $-18.03 \pm 0.15$ & $4776 \pm 50$ & $2.0241 \pm 0.0110$ & $-0.24 \pm 0.08$ & $10.98 \pm 0.50$ & $3.31^{+0.12}_{-0.06}$ & $29.05^{+0.34}_{-0.21}$ & $0.26^{+0.02}_{-0.02}$ \\
HD 189750 & $-62.65 \pm 0.06$ & $4814 \pm 50$ & $2.4569 \pm 0.0070$ & $-0.34 \pm 0.08$ & $4.15 \pm 0.50$ & $1.29^{+0.09}_{-0.09}$ & $11.01^{+0.29}_{-0.30}$ & $3.60^{+1.10}_{-0.70}$ \\
HD 226754 & $18.66 \pm 0.10$ & $4184 \pm 50$ & $1.7379 \pm 0.0110$ & $-0.12 \pm 0.08$ & $5.33 \pm 0.50$ & $1.31^{+0.12}_{-0.11}$ & $25.50^{+0.77}_{-0.79}$ & $4.40^{+1.60}_{-1.10}$
\enddata
\end{deluxetable}

\begin{deluxetable}{cccccccc}
\tablecaption{Chemical abundances relative to iron for stars in the red giant sample as determined by BACCHUS,  differential line-by-line comparison to Arcturus, as described in Section~\ref{spectroscopy}, for the elements Mg, Ti, Si, Ca, Al, V, and Ni. Dashes indicate elements for which abundances could not be reliably computed.The catalogue of abundances for neutron capture elements continues in Table~\ref{elems2}.\label{elems1}}
\tablehead{\colhead{Object} & \colhead{[Mg/Fe]} & \colhead{[Ti/Fe]} & \colhead{[Si/Fe]} & \colhead{[Ca/Fe]} & \colhead{[Al/Fe]} & \colhead{[V/Fe]} & \colhead{[Ni/Fe]}}
\startdata
BD+36 3564 & $0.38 \pm 0.10$ & $0.13 \pm 0.10$ & $0.23 \pm 0.02$ & $-0.05 \pm 0.00$ & $0.18 \pm 0.01$ & $0.00 \pm 0.00$ & $-0.03 \pm 0.04$ \\
BD+39 3577 & $0.25 \pm 0.03$ & $-0.10 \pm 0.04$ & $0.06 \pm 0.02$ & $0.04 \pm 0.03$ & $0.10 \pm 0.01$ & $-0.12 \pm 0.02$ & $-0.07 \pm 0.03$ \\
BD+42 3150 & $0.14 \pm 0.05$ & $0.10 \pm 0.05$ & $0.09 \pm 0.02$ & $0.01 \pm 0.01$ & $0.14 \pm 0.02$ & $0.17 \pm 0.02$ & $0.02 \pm 0.03$ \\
BD+43 3171 & -- & $-0.21 \pm 0.10$ & $0.22 \pm 0.21$ & $-0.24 \pm 0.04$ & $0.15 \pm 0.03$ & $-0.12 \pm 0.10$ & $-0.20 \pm 0.21$ \\
BD+48 2904 & $0.07 \pm 0.03$ & $0.07 \pm 0.03$ & $0.12 \pm 0.02$ & $0.05 \pm 0.07$ & $0.22 \pm 0.01$ & $0.15 \pm 0.02$ & $-0.01 \pm 0.04$ \\
BD+48 2955 & $0.24 \pm 0.04$ & $-0.04 \pm 0.10$ & $0.20 \pm 0.04$ & $-0.10 \pm 0.05$ & $0.12 \pm 0.10$ & $-0.04 \pm 0.04$ & $-0.08 \pm 0.05$ \\
HD 174020 & -- & $0.09 \pm 0.10$ & $0.02 \pm 0.14$ & $-0.03 \pm 0.06$ & $0.15 \pm 0.03$ & $0.12 \pm 0.10$ & $0.09 \pm 0.10$ \\
HD 174829 & $0.11 \pm 0.14$ & $0.16 \pm 0.04$ & $0.08 \pm 0.04$ & $-0.03 \pm 0.02$ & $0.14 \pm 0.02$ & $0.02 \pm 0.01$ & $-0.08 \pm 0.02$ \\
HD 175740 & -- & -- & -- & -- & -- & -- & -- \\
HD 175884 & $0.10 \pm 0.02$ & $0.18 \pm 0.03$ & $0.07 \pm 0.02$ & $-0.02 \pm 0.03$ & $0.14 \pm 0.01$ & $0.09 \pm 0.02$ & $-0.04 \pm 0.02$ \\
HD 178797 & $0.19 \pm 0.01$ & $0.10 \pm 0.02$ & $0.18 \pm 0.02$ & $-0.06 \pm 0.01$ & $0.16 \pm 0.02$ & $0.01 \pm 0.01$ & $-0.04 \pm 0.03$ \\
HD 178910 & $0.20 \pm 0.07$ & $0.13 \pm 0.05$ & $0.18 \pm 0.04$ & $0.09 \pm 0.01$ & $0.24 \pm 0.06$ & $0.36 \pm 0.06$ & $0.25 \pm 0.02$ \\
HD 179396 & $0.18 \pm 0.05$ & $0.09 \pm 0.02$ & $0.10 \pm 0.04$ & $-0.03 \pm 0.01$ & $0.17 \pm 0.03$ & $0.15 \pm 0.02$ & $-0.06 \pm 0.03$ \\
HD 180312 & $0.17 \pm 0.01$ & $0.18 \pm 0.07$ & $0.04 \pm 0.03$ & $-0.04 \pm 0.04$ & $0.18 \pm 0.00$ & $0.04 \pm 0.02$ & $-0.07 \pm 0.03$ \\
HD 180475 & $0.19 \pm 0.04$ & $0.23 \pm 0.07$ & $0.12 \pm 0.03$ & $-0.11 \pm 0.01$ & $0.08 \pm 0.01$ & $-0.08 \pm 0.01$ & $-0.07 \pm 0.02$ \\
HD 180658 & $0.13 \pm 0.05$ & $0.15 \pm 0.01$ & $0.05 \pm 0.03$ & $0.03 \pm 0.01$ & $0.20 \pm 0.02$ & $0.19 \pm 0.02$ & $-0.02 \pm 0.02$ \\
HD 180682 & -- & $0.31 \pm 0.01$ & $0.22 \pm 0.04$ & $0.05 \pm 0.05$ & $0.27 \pm 0.02$ & $0.12 \pm 0.03$ & $-0.01 \pm 0.05$ \\
HD 180682 & $0.38 \pm 0.01$ & $0.19 \pm 0.10$ & $0.31 \pm 0.02$ & $-0.15 \pm 0.02$ & $0.12 \pm 0.01$ & $-0.15 \pm 0.01$ & $-0.03 \pm 0.03$ \\
HD 181022 & -- & $0.04 \pm 0.10$ & $0.27 \pm 0.08$ & $-0.16 \pm 0.04$ & $0.26 \pm 0.08$ & $-0.01 \pm 0.03$ & $-0.24 \pm 0.15$ \\
HD 181069 & $0.04 \pm 0.06$ & $0.17 \pm 0.01$ & $0.06 \pm 0.04$ & $-0.01 \pm 0.00$ & $0.14 \pm 0.03$ & $0.18 \pm 0.01$ & $0.04 \pm 0.03$ \\
HD 181097 & $0.24 \pm 0.04$ & $0.12 \pm 0.02$ & $0.11 \pm 0.03$ & $0.04 \pm 0.04$ & $0.21 \pm 0.03$ & $0.14 \pm 0.02$ & $-0.08 \pm 0.03$ \\
HD 181597 & $0.08 \pm 0.01$ & $0.13 \pm 0.02$ & $0.01 \pm 0.02$ & $0.06 \pm 0.02$ & $0.14 \pm 0.01$ & $0.18 \pm 0.01$ & $0.00 \pm 0.02$ \\
HD 181778 & $0.04 \pm 0.01$ & $0.07 \pm 0.02$ & $0.01 \pm 0.02$ & $-0.01 \pm 0.00$ & $0.13 \pm 0.04$ & $0.10 \pm 0.01$ & $-0.04 \pm 0.02$ \\
HD 181880 & $0.26 \pm 0.11$ & $0.06 \pm 0.05$ & $0.20 \pm 0.03$ & $-0.05 \pm 0.01$ & $0.19 \pm 0.01$ & $0.04 \pm 0.03$ & $-0.04 \pm 0.04$ \\
HD 182354 & $0.04 \pm 0.09$ & $0.09 \pm 0.01$ & $0.05 \pm 0.02$ & $0.01 \pm 0.02$ & $0.05 \pm 0.07$ & $0.08 \pm 0.01$ & $-0.05 \pm 0.03$ \\
HD 182531 & $0.08 \pm 0.01$ & $0.14 \pm 0.10$ & $0.06 \pm 0.04$ & $-0.04 \pm 0.05$ & $0.13 \pm 0.02$ & $0.02 \pm 0.06$ & $0.01 \pm 0.02$ \\
HD 182692 & $0.10 \pm 0.05$ & $0.15 \pm 0.02$ & $0.08 \pm 0.03$ & $0.09 \pm 0.05$ & $0.22 \pm 0.04$ & $0.17 \pm 0.02$ & $-0.01 \pm 0.03$ \\
HD 182694 & $0.03 \pm 0.01$ & $0.14 \pm 0.04$ & $-0.00 \pm 0.02$ & $-0.00 \pm 0.01$ & $0.05 \pm 0.01$ & $0.04 \pm 0.03$ & $-0.13 \pm 0.01$ \\
HD 183124 & $0.22 \pm 0.05$ & $0.10 \pm 0.01$ & $0.08 \pm 0.03$ & $0.03 \pm 0.02$ & $0.17 \pm 0.01$ & $0.12 \pm 0.01$ & $-0.04 \pm 0.03$ \\
HD 185286 & $0.22 \pm 0.02$ & $-0.04 \pm 0.10$ & $0.10 \pm 0.03$ & $-0.00 \pm 0.04$ & $0.15 \pm 0.02$ & $0.17 \pm 0.06$ & $0.08 \pm 0.03$ \\
HD 188537 & $0.26 \pm 0.03$ & $-0.02 \pm 0.07$ & $0.12 \pm 0.02$ & $0.04 \pm 0.02$ & $0.19 \pm 0.10$ & $0.17 \pm 0.01$ & $-0.03 \pm 0.05$ \\
HD 189750 & $0.13 \pm 0.04$ & $0.15 \pm 0.04$ & $0.03 \pm 0.01$ & $0.01 \pm 0.02$ & $0.11 \pm 0.01$ & $0.11 \pm 0.02$ & $-0.04 \pm 0.01$ \\
HD 226754 & $0.25 \pm 0.03$ & $-0.01 \pm 0.10$ & $0.07 \pm 0.04$ & $-0.04 \pm 0.02$ & $0.25 \pm 0.06$ & $0.10 \pm 0.10$ & $-0.02 \pm 0.02$
\enddata
\end{deluxetable}

\begin{deluxetable}{ccccccc}
\tablecaption{Chemical abundances relative to iron of neutron capture elements for stars in the red giant sample as determined by BACCHUS, without differential line-by-line comparison to Arcturus, as described in Section~\ref{spectroscopy}, for the elements Sr, Y, Zr, Ba, La, and Eu. Dashes indicate elements for which abundances could not be reliably computed.\label{elems2}}
\tablehead{\colhead{Object} & \colhead{[Sr/Fe]} & \colhead{[Y/Fe]} & \colhead{[Zr/Fe]} & \colhead{[Ba/Fe]} & \colhead{[La/Fe]} & \colhead{[Eu/Fe]}}
\startdata
BD+36 3564 & $-0.13 \pm 0.11$ & $-0.45 \pm 0.01$ & $-0.24 \pm 0.04$ & $0.33 \pm 0.10$ & $0.07 \pm 0.05$ & -- \\
BD+39 3577 & $-0.19 \pm 0.10$ & $-0.30 \pm 0.05$ & $-0.24 \pm 0.11$ & $0.26 \pm 0.10$ & $-0.39 \pm 0.01$ & $-0.09 \pm 0.10$ \\
BD+42 3150 & $0.22 \pm 0.08$ & $-0.13 \pm 0.06$ & $0.05 \pm 0.03$ & $0.12 \pm 0.10$ & $0.09 \pm 0.04$ & $0.19 \pm 0.10$ \\
BD+43 3171 & $0.22 \pm 0.18$ & $-0.29 \pm 0.09$ & $-0.06 \pm 0.20$ & $-0.04 \pm 0.10$ & $0.08 \pm 0.23$ & -- \\
BD+48 2904 & $0.05 \pm 0.04$ & $-0.22 \pm 0.07$ & $0.05 \pm 0.04$ & $0.18 \pm 0.10$ & $0.16 \pm 0.06$ & -- \\
BD+48 2955 & $0.05 \pm 0.11$ & $-0.06 \pm 0.03$ & $-0.05 \pm 0.06$ & -- & $0.18 \pm 0.06$ & -- \\
HD 174020 & $0.05 \pm 0.09$ & $-0.08 \pm 0.17$ & $0.31 \pm 0.14$ & -- & $0.11 \pm 0.24$ & $0.07 \pm 0.10$ \\
HD 174829 & $-0.09 \pm 0.03$ & $-0.26 \pm 0.05$ & $-0.14 \pm 0.04$ & -- & $0.12 \pm 0.05$ & -- \\
HD 175740 & -- & -- & -- & -- & -- & -- \\
HD 175884 & $-0.14 \pm 0.02$ & $-0.22 \pm 0.05$ & $-0.08 \pm 0.04$ & -- & $0.16 \pm 0.06$ & -- \\
HD 178797 & $-0.23 \pm 0.10$ & $-0.24 \pm 0.07$ & $-0.19 \pm 0.04$ & $0.37 \pm 0.10$ & $0.04 \pm 0.05$ & -- \\
HD 178910 & $0.00 \pm 0.13$ & $-0.28 \pm 0.05$ & $-0.08 \pm 0.04$ & $0.29 \pm 0.10$ & $-0.14 \pm 0.07$ & -- \\
HD 179396 & $0.05 \pm 0.05$ & $-0.25 \pm 0.07$ & $-0.05 \pm 0.03$ & $0.32 \pm 0.10$ & $0.00 \pm 0.04$ & $-0.04 \pm 0.10$ \\
HD 180312 & $0.15 \pm 0.17$ & $-0.26 \pm 0.05$ & $0.01 \pm 0.05$ & $0.16 \pm 0.10$ & $0.10 \pm 0.06$ & -- \\
HD 180475 & $-0.06 \pm 0.20$ & $-0.28 \pm 0.09$ & $-0.27 \pm 0.03$ & $0.40 \pm 0.10$ & $0.11 \pm 0.04$ & -- \\
HD 180658 & $-0.03 \pm 0.01$ & $-0.35 \pm 0.06$ & $-0.01 \pm 0.06$ & $0.12 \pm 0.10$ & $-0.12 \pm 0.05$ & -- \\
HD 180682 & $0.06 \pm 0.22$ & $-0.28 \pm 0.26$ & $-0.13 \pm 0.18$ & $0.20 \pm 0.10$ & $0.01 \pm 0.14$ & -- \\
HD 180682 & $0.15 \pm 0.10$ & $-0.43 \pm 0.03$ & $-0.42 \pm 0.03$ & $0.16 \pm 0.10$ & $-0.09 \pm 0.04$ & -- \\
HD 181022 & $0.10 \pm 0.20$ & $-0.41 \pm 0.16$ & $0.19 \pm 0.17$ & $-0.04 \pm 0.10$ & $0.09 \pm 0.12$ & -- \\
HD 181069 & $0.09 \pm 0.09$ & $-0.09 \pm 0.09$ & $-0.12 \pm 0.05$ & $0.33 \pm 0.10$ & $0.01 \pm 0.04$ & $0.10 \pm 0.10$ \\
HD 181097 & $0.07 \pm 0.11$ & $-0.19 \pm 0.04$ & $0.01 \pm 0.04$ & -- & $0.14 \pm 0.04$ & -- \\
HD 181597 & $0.05 \pm 0.03$ & $-0.15 \pm 0.07$ & $0.03 \pm 0.04$ & $0.28 \pm 0.10$ & $0.12 \pm 0.06$ & $0.18 \pm 0.10$ \\
HD 181778 & $-0.06 \pm 0.08$ & $-0.19 \pm 0.06$ & $0.00 \pm 0.05$ & -- & $0.00 \pm 0.05$ & $0.14 \pm 0.10$ \\
HD 181880 & $-0.06 \pm 0.07$ & $-0.24 \pm 0.08$ & $-0.08 \pm 0.04$ & -- & $0.11 \pm 0.06$ & -- \\
HD 182354 & $-0.05 \pm 0.07$ & $-0.11 \pm 0.07$ & $0.00 \pm 0.03$ & $0.42 \pm 0.10$ & $0.12 \pm 0.05$ & -- \\
HD 182531 & $-0.16 \pm 0.07$ & $-0.22 \pm 0.06$ & $-0.03 \pm 0.04$ & -- & $0.05 \pm 0.06$ & -- \\
HD 182692 & $-0.15 \pm 0.12$ & $-0.30 \pm 0.08$ & $-0.06 \pm 0.05$ & $0.20 \pm 0.10$ & $-0.03 \pm 0.05$ & $0.06 \pm 0.10$ \\
HD 182694 & $-0.02 \pm 0.22$ & $-0.11 \pm 0.03$ & $0.09 \pm 0.05$ & $0.54 \pm 0.10$ & $0.14 \pm 0.04$ & $0.16 \pm 0.10$ \\
HD 183124 & $-0.02 \pm 0.30$ & $-0.25 \pm 0.03$ & $-0.07 \pm 0.07$ & $0.28 \pm 0.10$ & $0.05 \pm 0.03$ & -- \\
HD 185286 & $-0.00 \pm 0.07$ & $-0.15 \pm 0.07$ & $0.21 \pm 0.07$ & -- & $0.03 \pm 0.07$ & -- \\
HD 188537 & $-0.28 \pm 0.10$ & $-0.08 \pm 0.09$ & $0.17 \pm 0.03$ & $0.18 \pm 0.10$ & $0.21 \pm 0.05$ & $0.27 \pm 0.10$ \\
HD 189750 & $-0.43 \pm 0.10$ & $-0.17 \pm 0.03$ & $-0.00 \pm 0.05$ & $0.27 \pm 0.10$ & $0.07 \pm 0.03$ & $0.17 \pm 0.10$ \\
HD 226754 & $-0.00 \pm 0.07$ & $-0.43 \pm 0.08$ & $0.01 \pm 0.06$ & -- & $-0.04 \pm 0.07$ & --
\enddata
\end{deluxetable}

\section{Results}
\label{targets}

\subsection{Red Giants}
\label{rgs}

We determined the mass, radius, and age for the 33 red giants from their atmospheric and asteroseismic observables (see Table~\ref{astero_table}) using the BAyesian STellar Algorithm \citep[\textsc{basta}][]{silvaaguirre2015,silvaaguirre2017}. \textsc{basta} compares the observed properties (\teff, \feh, \logg, \Dnu, and \numax) with predictions from theoretical models of stellar evolution, in this case the recently updated BaSTI (a Bag of Stellar Tracks and Isochrones) stellar models and isochrones library \citep{Hidalgo:2018dy}. The isochrones include core overshooting with an efficiency of 0.20 times the pressure scale height as described in \citet{Hidalgo:2018dy}, but do not include diffusion or rotational mixing.

The spectroscopic properties are the effective temperature \teff, the metallicity \feh, and the surface gravities \logg from Table~\ref{stellar_props}. These are accompanied by the global asteroseismic properties \Dnu and \numax from Table~\ref{astero_table}. Theoretical predictions of \Dnu and \numax were computed using the asteroseismic scaling relation for any point along an evolutionary track or isochrone. For the solar values, we adopted $\nu_{\text{max, }\odot}=3090$~$\mu$Hz, $\Delta\nu_{\odot}=135.1$~$\mu$Hz \citep{huber2011}, and $T_{\text{eff, }\odot}=5777$~K. We emphasize that all quoted error bars in Tables~\ref{stellar_props} and~\ref{astero_table} are formal uncertainties, and do not take into account systematic differences in asteroseismic measurement methods, effective temperature scales, or stellar model physics. For example, recent tests of red giant models imply that systematic age errors can be expected to be significantly larger than the formal age uncertainties in Table~\ref{stellar_props} \citep{Tayar2017}.

The accuracy of the asteroseismic scaling relations across different metallicities, effective temperatures, and evolutionary status is currently under discussion \citep[see][]{white2011,belkacem2011,sharma2016,viani2017}.
We applied the correction by \cite{Serenelli:2017cn} to the large frequency separation relation in Equation~\ref{Dnuscaling} as it has been shown to reproduce the results of a number of classical age determination for the open clusters M67 \citep{stello2016} and NGC~6819 \citep{casagrande2016}.

\begin{figure}
\plotone{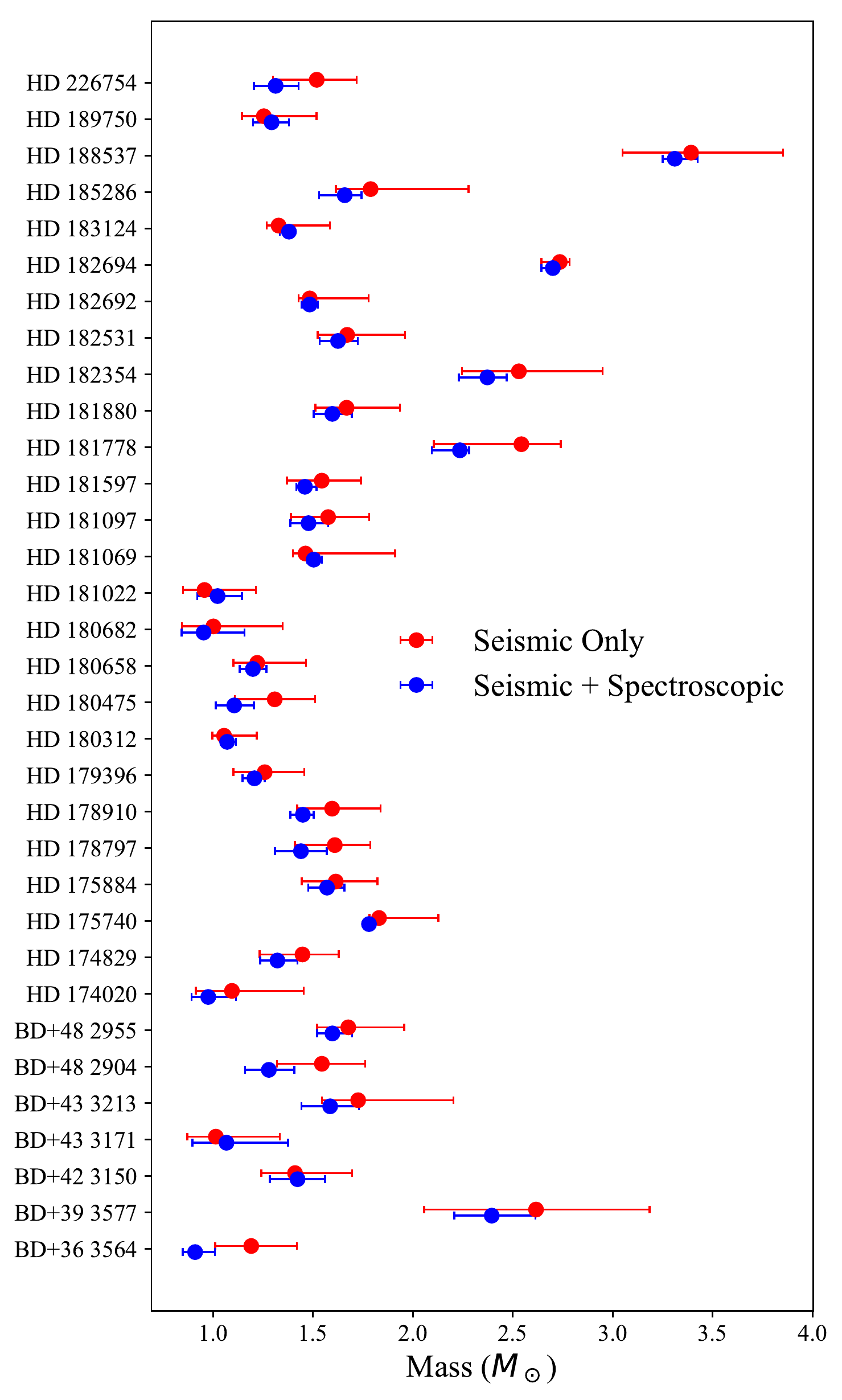}
\caption{\label{masses}
Using asteroseismic constraints only (red) and asteroseismic and spectroscopic constraints jointly (blue) we infer the masses of each star in the asteroseismic sample of giants.}
\end{figure}

We compare the solutions found using this set of fitting parameters with those found using only asteroseismic input in Figure~\ref{masses}, and we find that the change in median stellar parameters between the two results is small for all analyzed red giants. However, in the comparison plot it becomes clear that adding the spectroscopic constraints to the fit reduces the posterior uncertainty in stellar mass.

\begin{figure}
\plotone{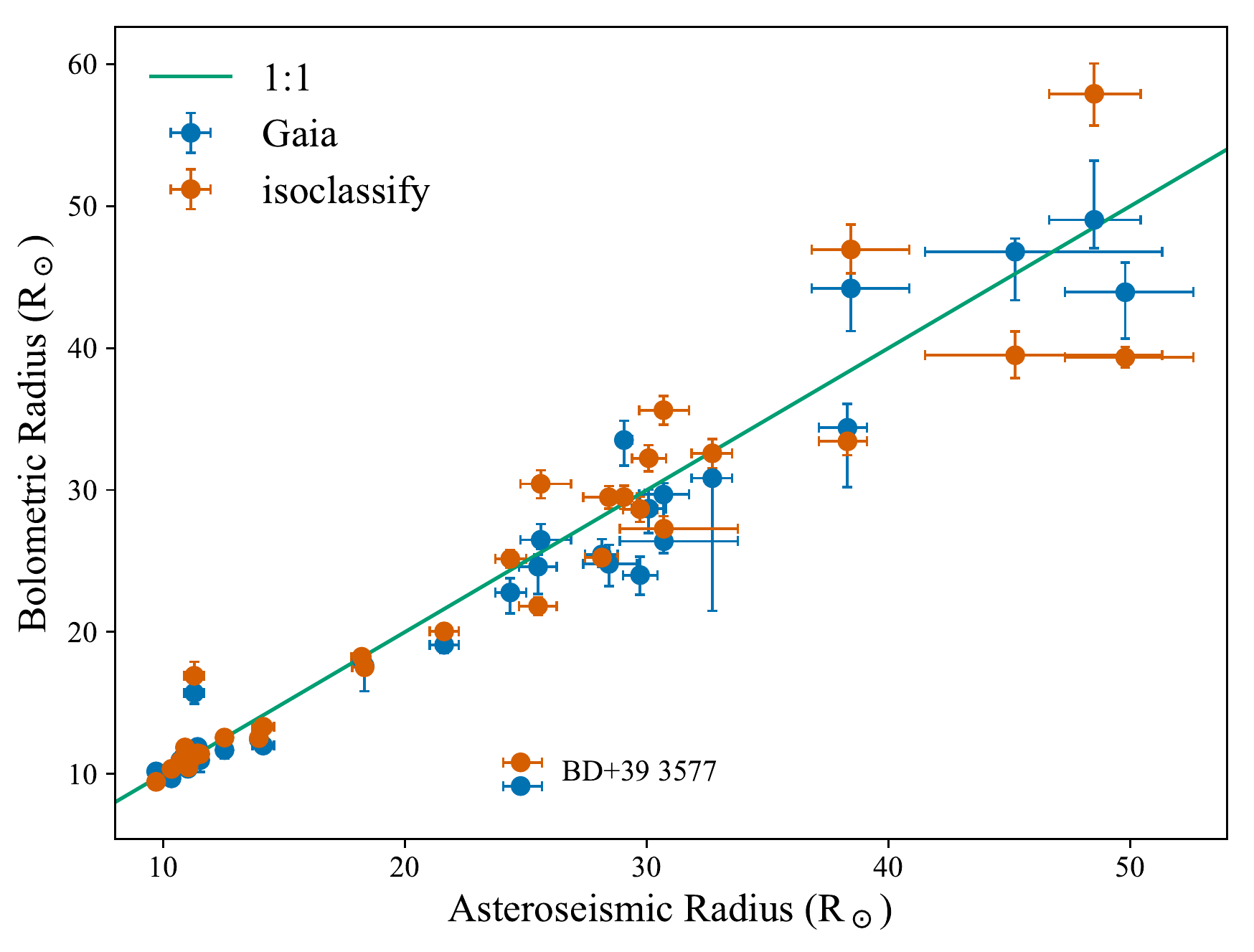}

\caption{\label{gaia_comparison}
A comparison of the stellar radii determined here from asteroseismology and spectroscopy to those from the \gaia~DR2 catalogue ($y$-axis) and from our own calculations based on \gaia parallaxes and TRES spectroscopy ($x$-axis). Blue points are from \gaia measurements explicitly, and orange points \gaia-like calculations using \texttt{isoclassify} \citep{huber2017} and substituting own $T_{eff}$ measurements rather than those from \gaia. The green line overlaid shows a 1:1 relation. There is overall good agreement except for very large radii, and except for BD+39~3577 (marked).}
\end{figure}

To gauge the level of improvement in our understanding of these stars, in Figure~\ref{gaia_comparison} we plot the radii determined here against those from the \gaia~DR2 catalog determined by the stellar bolometric flux and parallax. 
We also calculate \gaia-like stellar bolometric radii using the software \texttt{isoclassify} \citep{huber2017}, using our new spectroscopic \teff, \feh, and \logg measurements rather than those from \gaia~DR2, together with \gaia~DR2 parallaxes, and SIMBAD $V$. While not all of our targets have radii in the \gaia catalogs, for those that do we find that there is overall agreement to within a few $\sigma$, but that the results from stellar modelling are consistently slightly larger than those from \gaia. 
This discrepancy goes away for most stars when we use our own \texttt{isoclassify} radii, suggesting that this is an effect of \teff calibration. Two stars are noticeably very different: BD+39~3577 has a precise \gaia radius of $9.14^{+0.25}_{-0.13}$~\rsun, but $24.78^{+0.88}_{-0.72}$~\rsun from modelling. It is unclear why this would be the case; it is possible that an unidentified binary companion has affected either the asteroseismic detection or parallax. Likewise BD+42~3150 has a \gaia radius of $15.70^{+0.52}_{-0.76}$~\rsun but $11.27^{+0.39}_{-0.41}$~\rsun from stellar modelling, but of our sample it has the lowest \gaia parallax over error at only 18.0, and this anomalously high value is likely due to noise. Agreement for large-radius ($>30 \rsun$) stars is somewhat poorer, even though the \gaia parallax-over-error is apparently adequate. This dispersion, unlike for medium radius stars, is more pronounced for the \texttt{isoclassify} radii than for \gaia~DR2, suggesting that our $V$ band bolometric corrections for these more yellow (\gaia $Bp-Rp \sim 0.25$) stars are insufficiently accurate. 

\subsubsection{Chemical Compositions}
\label{chemical}
\begin{figure*}
\plotone{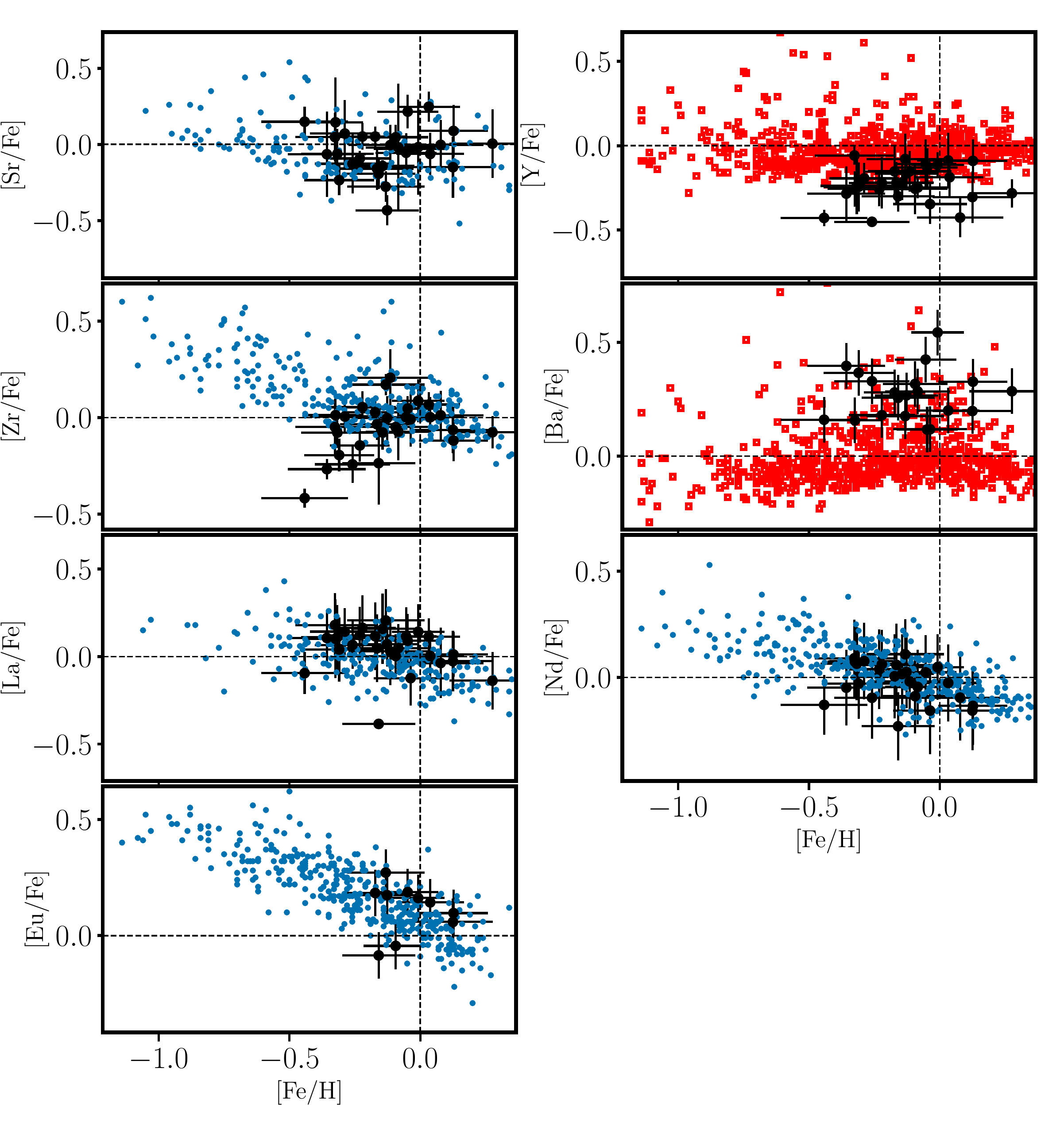}
\caption{\label{fig:ncap}
The [Mg/Fe], [Si/Fe], [Ti/Fe], [V/Fe] (left panel) and [Al/Fe], [Ca/Fe], [Ni/Fe], [Cr/Fe] (right panel) abundance ratios as a function of iron for our stars (black circles). We also show a representative sample of Galactic disk stars from \citet[red,][]{Bensby2014}, \citet[blue][]{Battistini2016}. These elemental ratios give a representive example of the chemical composition of our sample and show they are consistent with the Galactic disk population. } 
\end{figure*}

The chemical composition for each star was measured in the $\alpha$ (Mg, Ti, Si, Ca), odd-Z (Al, V) and Fe-peak (Fe, Ni) elemental families in a differential way with respect to Arcturus. The chemical composition for the neutron-capture elements are shown in Fig.~\ref{fig:ncap}, and were derived in absolute terms rather than differentially with respect to Arcturus. The elemental abundance ratios were measured in order to determine the Galactic populations to which these stars belong. The metallicities, which are tabulated in Table~\ref{stellar_props}, are too high (with $-0.51<$~[M/H]~$<$ +0.14~dex) to belong to the Galactic halo, whose peak metallicity is around $\sim -1.50$ \cite[e.g.][]{Chiba2000}. Furthermore, the distance distribution (Table~\ref{all_stars}) indicates that all stars are located within a few~kpc of the Sun and are not part of the Galactic bulge. Thus, these stars are drawn from only the Galactic thick and thin disks. We provide a detailed chemical abundance analysis below to support this claim.\\

\begin{figure*}
\plotone{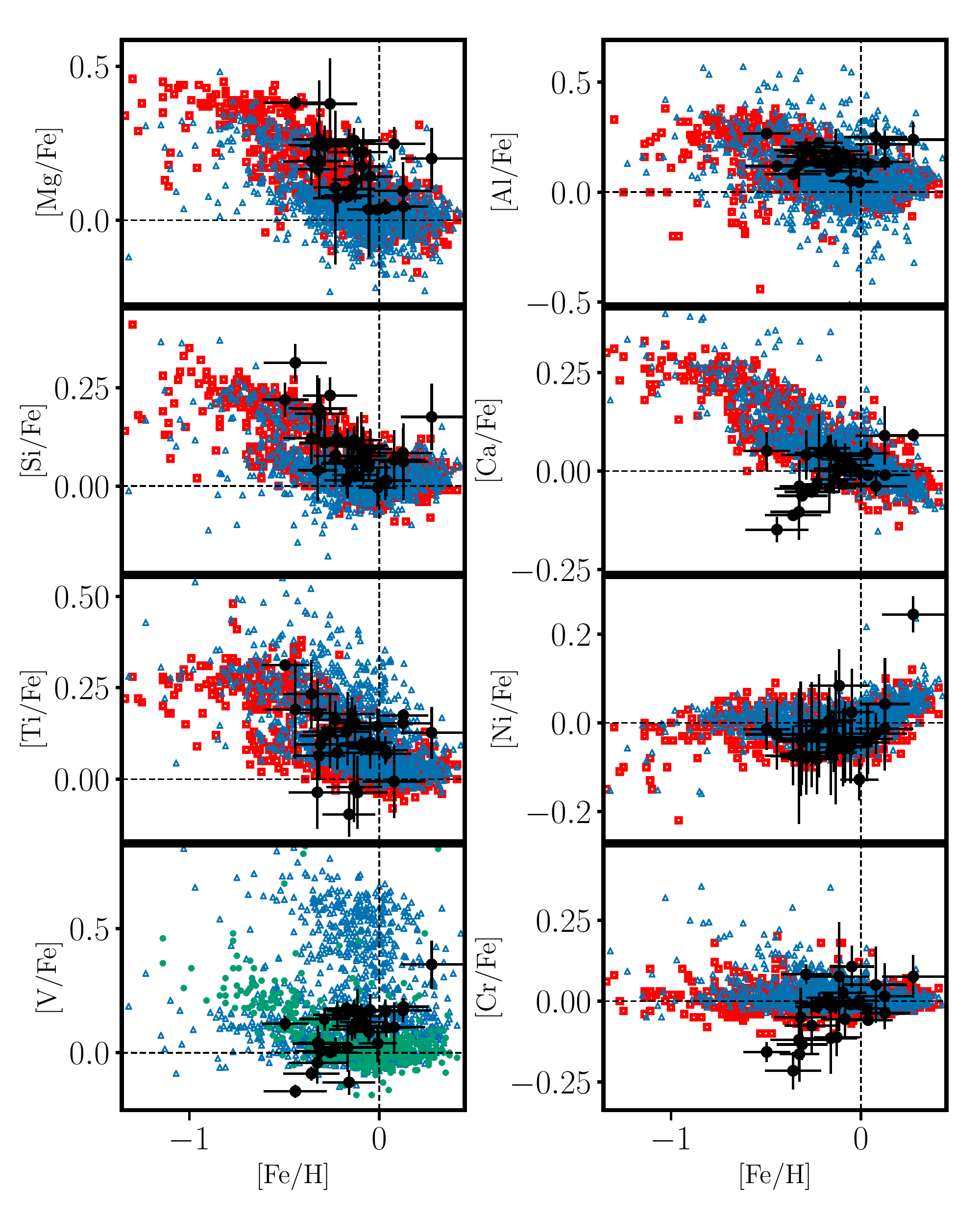}

\caption{\label{alphael}
The [Mg/Fe], [Si/Fe], [Ti/Fe], [V/Fe] (left panel) and [Al/Fe], [Ca/Fe], [Ni/Fe], [Cr/Fe] (right panel) abundance ratios as a function of iron for our stars (black circles). We also show a representative sample of Galactic disk stars from \citet[open red squares]{Bensby2014}, \citet[open blue squares]{Adibekyan2012}, and \citet[teal circles]{Battistini2015}. These elemental ratios show that the chemical composition of our sample is consistent with the Galactic disk population. } 
\end{figure*}

One of the primary ways to determine, in a chemical sense, whether the stars in our sample are drawn from the Galactic disk(s), bulge or halo, is with the ratio of their $\alpha$-elements to Fe. The $\alpha$ elements are formed after He burning (e.g. Mg, Ti, Si, Ca) and largely dispersed into the interstellar medium through Type~II supernovae (SNII) \citep{Matteucci2001}. 
The Galactic disk can be chemically dissected into a low- and a high-alpha component that have different vertical and age structure \citep[see e.g.][]{2016ApJ...823...30B,2017A&A...608L...1H,2018MNRAS.475.5487S}, and are commonly associated with the thin and the thick disk \citep[e.g.][and references therein]{Edvardsson1993, Adibekyan2012, Feltzing2013, Bensby2014}. At a given metallicity, the thick disk is enhanced in [Mg, Si, Ca, Ti/Fe] compared to the Galactic thin disk.

In Fig.~\ref{alphael}, we display the [Mg/Fe] abundance ratio as a function of [Fe/H] for our stars (black circles), compared\footnote{There may be systematics between our [X/Fe] abundance scale and those of our comparison samples.} to representative thick and thin disk stars from \cite[open red squares]{Bensby2014} and \cite[open blue triangles]{Adibekyan2012}. 

For most of the stars in our sample, the [Mg/Fe] abundance ratios are enhanced. This is true for all of the $\alpha$-elements except Ca where the is a much larger spread. The commonly used [$\alpha$/Fe] abundance ratio, the average of Mg, Ti, Si, Ca (thus it is ([Mg/Fe] + [Ca/Fe] + [Si/Fe] + [Ti/Fe] / 4) ), is also enhanced in most stars. This is consistent with most of the stars observed here belonging to the Galactic disks with a slight (of order $\sim$0.15~dex) enhancement in the $\alpha$-elements. Fig.~\ref{alphael} clearly rule out the Galactic bulge (which would require the sample to be significantly more $\alpha$-enriched given their metallicity) and the Galactic halo (given that the stars would need to be significantly more metal-poor). 

In addition, to the $\alpha$ and odd-Z elements we also derived the chemical abundance for several neutron capture elements including Sr, Zr, La, Eu (left panel of Fig~\ref{fig:ncap}) as well as Y, Ba, and Nd (right panel of Fig.~\ref{fig:ncap}).  It is clear from Fig.~\ref{fig:ncap}  that the chemical abundance ratio of each neutron capture element is consistent with the Galactic disk population. The Ba of our sample is slightly enhanced, however, while the Y of our sample is slightly reduced relative to the general disk population of \cite{Bensby2014}. Nevertheless, we conclude all elemental abundance ratios studied our sample most closely resemble the Galactic disk. 

We note that one of the stars (HD~175740) can also be found in the Hypatia catalogue \citep{hypatia}. The chemical abundance ratios in each element are consistent, within the uncertainties (of order $\sim$0.10--0.15~dex for most elemental abundance ratios in Hypatia and up to $\sim$0.05~dex here). 

\subsubsection{Red Clump Stars}
\label{clumpstars}

Red clump stars, which burn helium in their cores, can be distinguished from hydrogen-shell burning giants asteroseismologically, via their much higher \textit{g}-mode period spacings \citep{rggmodehelium}. 
The term `red clump' arises from the fact that such stars can have a very narrow range of luminosities, so that they appear as a clump in the HR diagram \citep{2016ARA&A..54...95G}. This property makes them useful standard candles to which distances can be accurately computed from photometry. Red clump stars have been used to calibrate the \gaia survey's parallaxes at long distances \citep{2017A&A...598L...4D,2017MNRAS.471..722H,2018A&A...609A.116R}. \gaia~DR2 parallaxes have a zero-point offset of $\sim 0.03$~mas \citep{gaiadr2parallax}, and hierarchical models of the ensemble of \gaia clump stars can be used to accurately estimate this and thereby improve the accuracy of \gaia distances greater than a few~kpc (Hawkins et al., in prep.).

\begin{figure*}
\plotone{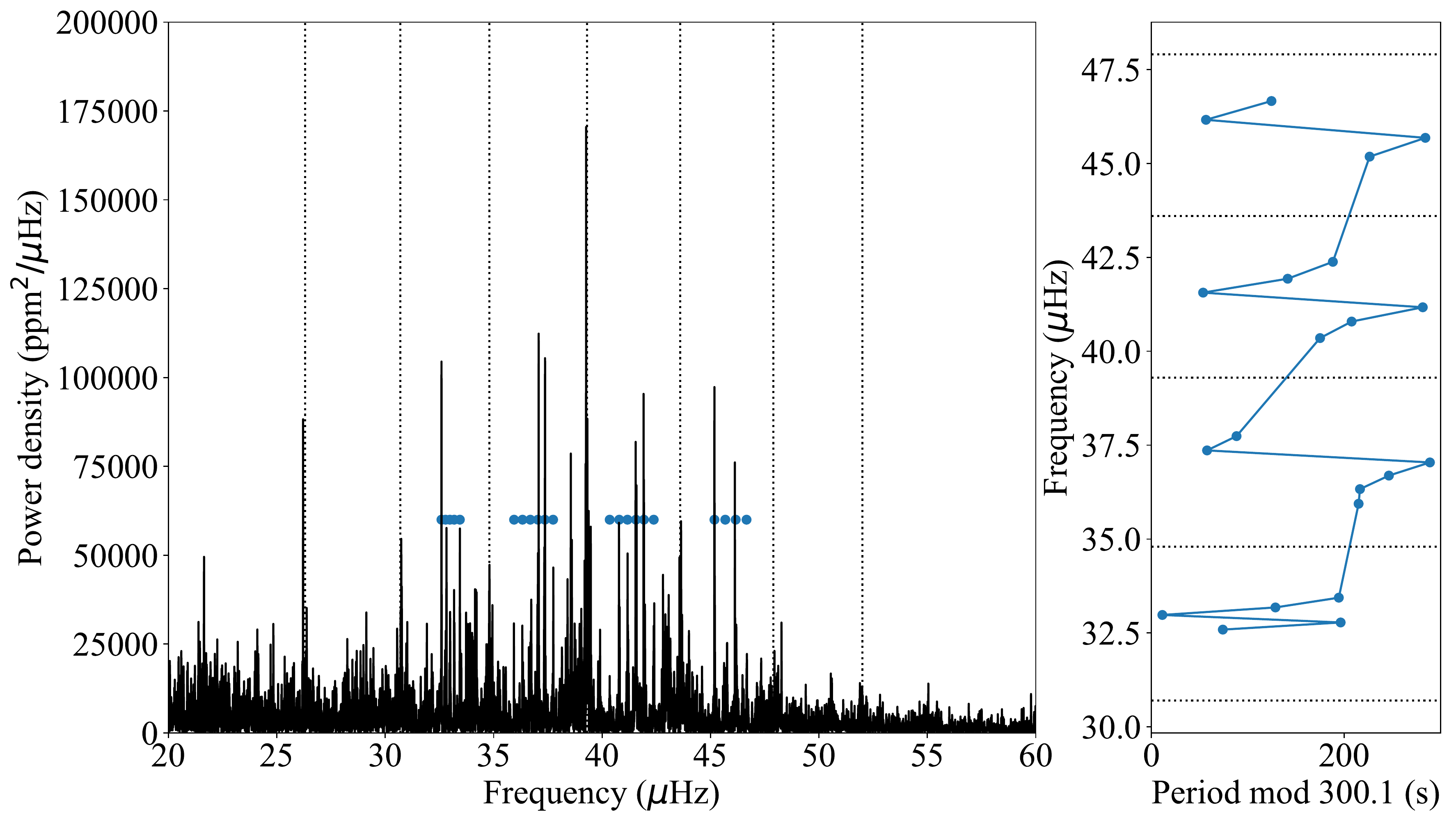}
\caption{\label{HD_183124}
Power spectrum (left) and period \'{e}chelle diagram (right) of the solar-like oscillations of the red clump star HD~183124. The modes in the power spectrum used for the period \'{e}chelle diagram are highlighted with blue dots. In the period \'{e}chelle diagram we see the characteristic pattern of `bumped' modes from avoided crossings between the comb of \textit{p}-modes and \textit{g}-mode oscillations with a period spacing of $\Delta \Pi = 300.1~\text{s}$.}
\end{figure*}

From inspection of the power spectra, HD~181069, HD~183124, HD~182354, HD~182692, and HD~180658 are seen to be red clump stars. A power spectrum of the best example of these, HD~183124, together with a period \'{e}chelle diagram used to estimate its \textit{g}-mode period spacing, are shown in Figure~\ref{HD_183124}. While precise characterization of these stars is beyond the scope of this paper, they are ideal candidates for anchoring models of the mass and metallicity dependence of red clump properties for calibrating \gaia and other distance measures. 

\subsection{Main-Sequence Stars}
\label{mainseq}

For all the main-sequence stars in our sample, we inspected light curves and power spectra to determine their variability class. In the following subsections, we will briefly comment on some of the findings. Since main-sequence variables are diverse, and the relevant scientific questions varied, we have attempted only a very preliminary study of these stars in this paper, leaving detailed analysis to future work.

Our sample includes pulsating stars of spectral types~B,~A, and~F, as listed in Table~\ref{all_stars}. 

The sample includes 5 $\delta$\,Sct stars, which show p-mode pulsation. These oscillation modes have particularly long lifetimes and stable frequencies, making them precise stellar clocks with periods of $\sim$2\,hr. These can be used to search for binarity and to obtain orbital parameters from photometry alone \citep{shibahashi&kurtz2012}. We used the phase-modulation (PM) method of \citet{murphyetal2014} to investigate whether any of these $\delta$\,Sct stars are binaries. Any phase modulation is converted into a light arrival-time (R\o mer) delay, and for a binary, the time delays of each mode should vary in unison. Nearly 350 PM binaries are known in the full \textit{Kepler} dataset \citep{murphyetal2018}.

\begin{figure*}
\plotone{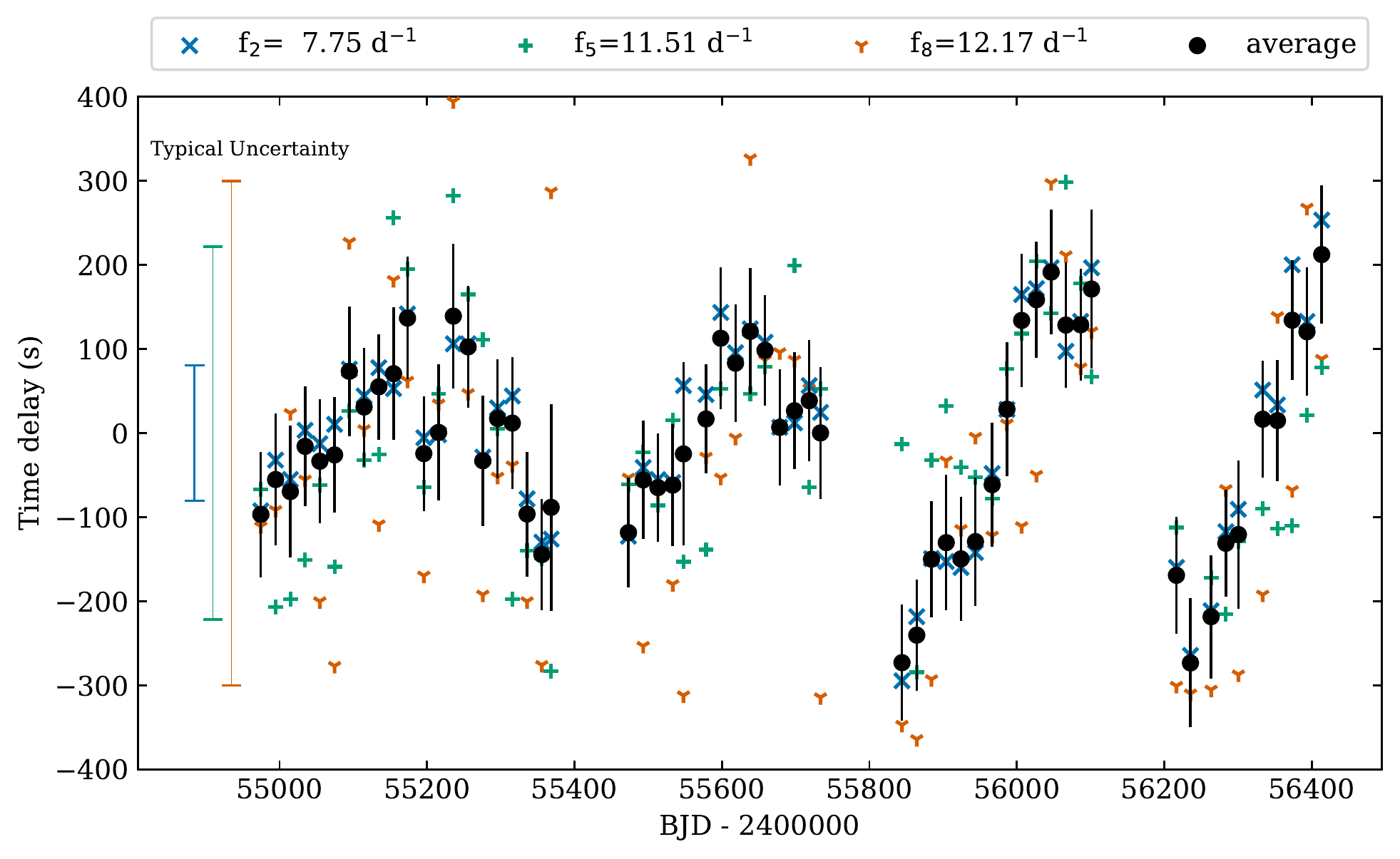}
\caption{\label{pm}
Time delay as a function of time for the $\delta$\,Sct star HD~186255, calculated for three of its oscillation modes ($f_2$, $f_5$, and $f_8$) and their average. These vary in unison, but with a small amplitude, and are possibly indicative of a low-mass or highly inclined companion.}
\end{figure*}

In four of the five targets we found evidence for binarity, while in the fifth (HD\,185397) there was some time-delay variation but there was no agreement between different modes so it is not of binary origin. Of the others, HD\,175841 and HD\,177781 are probably very long-period binaries, with periods far exceeding the Kepler datasets of $\sim$1470\,d. HD\,181521 appears to be an eccentric binary with a period of at least 1000\,d, but there is only 1 maximum and 1 minimum in the time-delay curve \citep[cf.][]{murphy&shibahashi2015}, so a unique orbital solution was unobtainable. Finally, HD\,186255 is probably a binary with a period of $\sim$415\,d (Figure~\ref{pm}), but there is a slight aperiodicity in the time delays, likely caused by beating between pulsation modes that are not well-separated in the frequency. That, coupled with the fact that this star falls on the failed Module~3 and is therefore missing data every 4th quarter (i.e. $\sim$93 of every 372.5\,d), makes the binary classification uncertain. If this is indeed a 415-d binary, the time delays are consistent with a companion of minimum mass $\sim$0.45\,M$_{\odot}$ in an orbit of moderate eccentricity ($\sim$0.15).

Several stars have a more complex classification than can be adequately noted in Table~\ref{all_stars}: HD~189684 is listed as an ellipsoidal variable, but also shows evidence for $\gamma$\,Dor variability. HD~185397 and HD~186255 are listed as $\gamma$\,Dor/$\delta$\,Sct hybrids, but may in fact simply be $\delta$\,Sct variables with nonlinear combination frequencies, and a detailed frequency analysis will be required to distinguish between these possibilities. HD~184788 shows a combination of two rotational modulation signals with base frequencies: 0.0885 and~$0.1966 c/d$. HD~184875 is a $\gamma$\,Dor but also shows evidence for an unknown contaminant. V554~Lyr and V2079~Cyg are both known $\alpha^2$\,CVn variables, which are chemically peculiar stars with strong magnetic fields that show rotational modulation. V2079~Cyg also shows a weak $\delta$\,Sct signal. The detection of rotational modulation in the chemically-peculiar HD~175132 suggests its reclassification as an $\alpha^2$\,CVn variable. 

There are two stars whose variability we classify as $\alpha^2$\,CVn, namely HD~176582 (B5V) and HD~179395 (B9), but which are not previously known to be chemically-peculiar. They have very short periods (1.58~d and 1.83~d respectively) and phase stability throughout the \kepler observations. While HD~176582 is listed as an eruptive variable by \citet{2016ApJ...829...23D}, this appears to be a misclassification considering the full \kepler smear light curve. Both stars show periods shorter than the shortest `heartbeat' binaries with tidally-induced pulsations from \citet{2012ApJ...753...86T}. Moreover, the variability periods are short enough that for a binary origin we would expect orbits to be circularized \citep{2000A&A...354..881D}. We suggest that these are nevertheless $\alpha^2$\,CVn variables, and that it will be valuable to study these stars spectroscopically for signs of chemical peculiarity.

The coherent g-mode pulsations in samples of B, A, and F stars observed by \kepler previously showed these stars to be near-rigid rotators \citep{kurtz2014,saio15,triana15,vanreeth15,vanreeth16,vanreeth18,murphy16,schmid16,moravveji16,ouazzani17,papics17,aerts17,szewczuk18,2018arXiv180907779A,2018A&A...618A..47C,2019MNRAS.482.1757L}. These studies cover about 100 stars so far and many more are to come. For 37 of the F-type pulsators among those, asteroseismic modelling of the g modes led to their masses, ages, core rotation and core mass with relative precisions of about 10\% \citep{2019MNRAS.tmp..495M}. However, the vast majority of intermediate-mass stars observed by Kepler have yet to be subjected to in-depth asteroseismic analyses and modelling of their interior properties. One of the valuable outputs of our current work includes the reduced light curves of several early-B stars, which were only scarcely targeted in the nominal \kepler mission. The few that were monitored did not reveal suitable oscillation frequency patterns to achieve a unique mode identification, which is a requirement to perform asteroseismic modelling. The investigation of pulsation modes in high-mass stars using high-quality \kepler smear data combined with high-precision spectroscopy to identify the modes \citep[Chapter 6]{aertsbook} is an exciting prospect for asteroseismology, as the interior physics of these stars are largely unknown \citep[e.g.][]{2019arXiv190502120B}, yet they play a pivotal role in stellar and galactic evolution. The in-depth asteroseismic analysis of the smear data for the~B~stars will be the subject of future work.

\subsubsection{Hump and Spike Stars}
\label{hs}

Several stars in the sample show the `hump-and-spike' morphology in their power spectra \citep[a broad `hump' of low-amplitude oscillations dominated by one high amplitude coherent oscillation toward the high frequency end of this band;][]{2013MNRAS.431.2240B,2014MNRAS.441.3543B,2017MNRAS.467.1830B}. \citet{2018MNRAS.474.2774S} have recently interpreted the hump-and-spike power spectra as evidence for Rossby modes.
These stars are marked `H+S' in Table~\ref{stellar_props}. Of these, HD~186155 and 14~Cyg are the third-- and sixth--brightest stars on silicon, making these the brightest stars that show this effect. The identification for HD~189178 is tentative, as the power spectrum also shows evidence of SPB pulsations. This is likewise the case for HD~183362 which shows $\gamma$\,Dor pulsations, and for HD~184787 there is long-term variability consistent with contamination. The other hump-and-spike identifications seem secure. The F5 star HD~186155, identified by SIMBAD as having a giant spectral type of F5II-III, is shown by its \gaia distance to in fact lie on the main-sequence. A detailed study of these stars will be presented by Antoci et al., in prep.

Another star with a hump-and-spike spectrum is Boyajian's~Star (KIC~8462852), which shows deep enigmatic dips in brightness \citep{2016MNRAS.457.3988B}, and has faded both throughout the \kepler mission \citep{2016ApJ...830L..39M} and in relation to Harvard photographic plates from 1890 onwards \citep{2016ApJ...822L..34S}. The dimming, which is chromatic in the manner expected of heterogeneous clouds of circumstellar dust in the line of sight \citep{2018ApJ...853..130D,2018arXiv180608842B}, has been ascribed to various causes \citep[reviewed in][]{2018RNAAS...2a..16W}, most notably a cloud of exocomets surrounding the star \citep[e.g.][]{2018MNRAS.473.5286W}. It is unclear whether the explanation of the hump-and-spike phenomenon will shed light on the strange behaviour of Boyajian's~Star, but it may be relevant.

\begin{figure}
\plotone{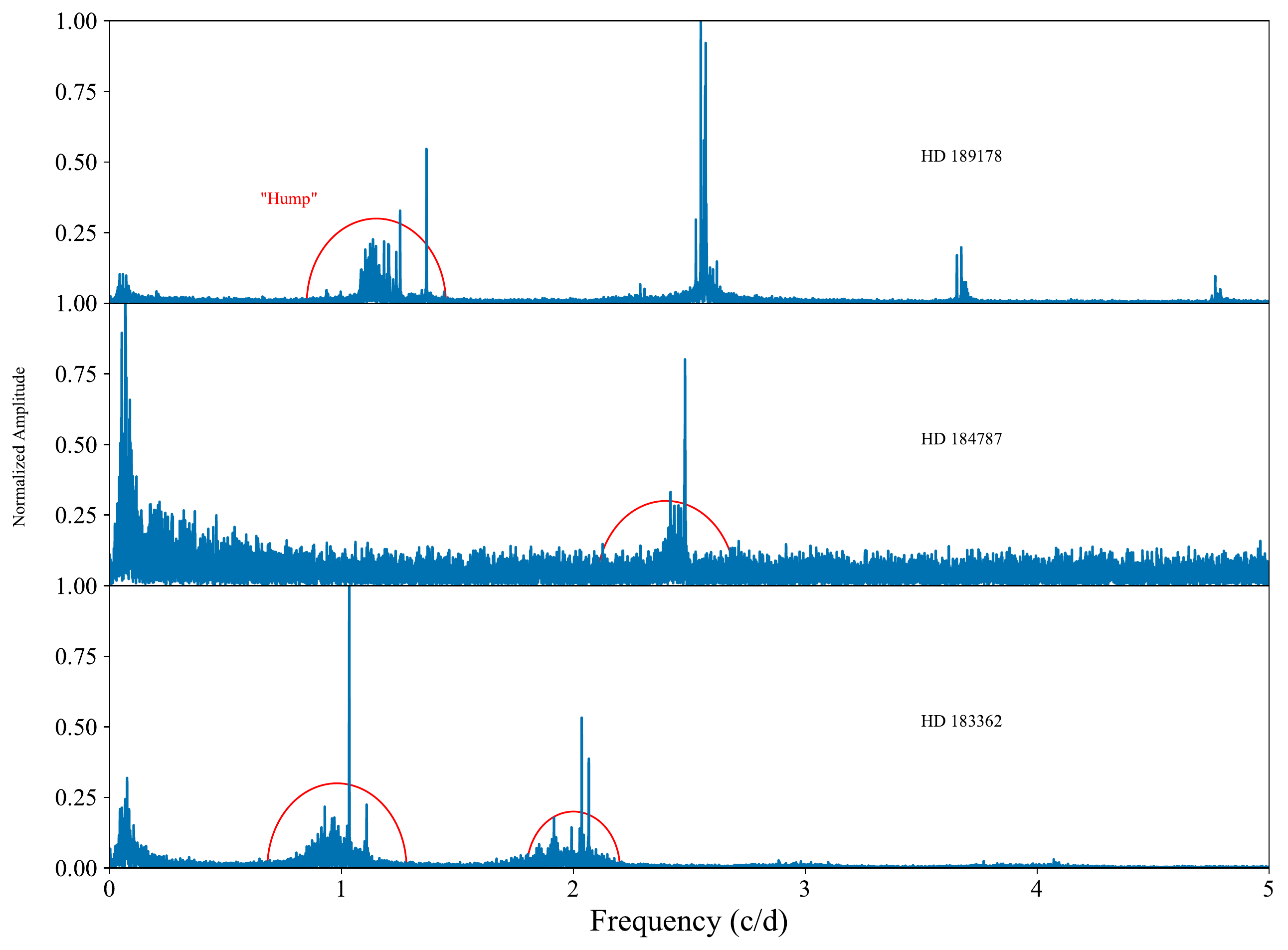}

\caption{\label{hs_fig}
Lomb-Scargle periodograms in normalized amplitude of three `Hump and Spike' stars: HD~189718, HD~184787, and HD~183362, with the `hump' features highlighted with red ellipses.}
\end{figure}

\subsubsection{Eclipsing Binaries}
\label{ebs}

We detect BD+47~2825 as a new eclipsing binary system, and obtained light curves for the previously-known eclipsing binaries HD~186994 \citep{2016AJ....151..101A}, V2083~Cyg \citep{2012MNRAS.421.1196Z}, and V380~Cyg \citep{2003A&A...399.1115C}. The known spectroscopic binary system HD~189684 \citep{2008MNRAS.389..869E} is newly identified as showing ellipsoidal variability, but does not show evidence of eclipses. We do not attempt detailed analysis of their variability in this paper.

\section{Open Science}
\label{open}

To facilitate open science, we have made the products of this research available online. All code used to produce smear light curves is available under a GPL~v3 license\footnote{ \href{https://github.com/benjaminpope/keplersmear}{\nolinkurl{github.com/benjaminpope/keplersmear}} \citep{keplersmear_zenodo}}. All smear light curves, both including the red giant sample studied in detail in Section~\ref{rgs}, and main-sequence stars as discussed in Sections~\ref{mainseq} and~\ref{ebs}, can be downloaded from the Mikulski Archive for Space Telescopes (MAST) as a High Level Science Product\footnote{\dataset[10.17909/t9-4sgf-9c19]{\doi{10.17909/t9-4sgf-9c19}}}. TRES spectra are available from the ExoFOP-TESS website\footnote{\dataset[exofop.ipac.caltech.edu/tess/]{https://exofop.ipac.caltech.edu/tess/}}.

All smear light curves in this paper, as well as the \LaTeX{} source code used to produce this document, can be found
at \href{https://github.com/benjaminpope/smearcampaign}{\nolinkurl{github.com/benjaminpope/smearcampaign}}\footnote{\dataset[10.5281/zenodo.3066218]{\doi{10.5281/zenodo.3066218}} \citep{smearcampaign_zenodo}}.

\section{Conclusions}
\label{conclusions}

The \kepler Smear Campaign establishes a legacy sample of~102 very bright stars, with \kepler light curves that in almost all cases reveal astrophysically interesting variability. The virtue of these bright stars is that they can be studied with interferometry, and more easily with spectroscopy than fainter targets, permitting especially detailed characterization. These stars will also be bright enough to be re-observed with high precision by the Transiting Exoplanet Survey Satellite \citep[TESS;][]{tess}. We have obtained detailed abundances of a subset of the red giants in this sample, with a view to confirming their membership of the Galactic thick and thin disk populations. A compelling next step is to use interferometric diameter measurements and to further constrain the red giant parameters, and compare these to the constraints from \gaia. Any tension between these measurements will help test and refine the asteroseismic scaling relations, and better models will propagate through to smaller systematic uncertainties in large samples of stars too faint for interferometry. Further improvements will be revealed by the detailed modelling of individual oscillation frequencies in these giants to infer interior structure such as convective overshoot. For the lower-frequency M~giants classed as LPVs in this paper, extending the systematics correction and quarter-stitching algorithms to more robustly correct their light curves without removing real signal will allow similar asteroseismic analysis, for a sample of stars that are much less well understood than their higher-frequency counterparts. 

The \kepler Smear Campaign has another natural extension: while many saturated stars in \ktwo have now been observed with `halo' apertures including their unsaturated pixels, many were not, either because they were fainter than the typical $Kp \lesssim 6.5$ limit, or because in Campaigns~0-3 and~5 no such apertures were selected. There is therefore the potential for a \ktwo Smear Campaign to complete the \ktwo sample down to fainter magnitudes, complementing the very brightest stars studied with halo photometry.

\section*{Acknowledgements} 

We would like to thank the anonymous reviewer, whose comments significantly improved this paper.

This work was performed in part under contract with the Jet Propulsion Laboratory (JPL) funded by NASA through the Sagan Fellowship Program executed by the NASA Exoplanet Science Institute. BJSP also acknowledges support from Balliol College and the Clarendon Fund. DH acknowledges support by the National Aeronautics and Space Administration through the K2 Guest Observer Program (NNX17AF76G, 80NSSC19K0108, 80NSSC18K0362) and the National Science Foundation (AST-1717000). TRW acknowledges the support of the Australian Research Council (grant DP150100250). TRW and VSA acknowledge
the support of the Villum Foundation (research grant 10118). DWL acknowledges partial support from the Kepler Extended Mission under NASA Cooperative Agreement NNX13AB58A with the Smithsonian Astrophysical Observatory. The research leading to these results has received funding from the European Research Council (ERC) under the European Union's Horizon 2020 research and innovation programme (grant agreement N$^\circ$670519: MAMSIE). VSA acknowledges support the Independent Research Fund Denmark (Research grant 7027-00096B). Funding for the Stellar Astrophysics Centre is provided by The Danish National Research Foundation (Grant agreement No.~DNRF106). PJ acknowledges FONDECYT Iniciaci\'on grant 11170174. 

BJSP acknowledges being on the traditional territory of the Lenape Nations and recognizes that Manhattan continues to be the home to many Algonkian peoples. We give blessings and thanks to the Lenape people and Lenape Nations in recognition that we are carrying out this work on their indigenous homelands. We would like to acknowledge the Gadigal Clan of the Eora Nation as the traditional owners of the land on which the University of Sydney is built and on which some of this work was carried out, and pay their respects to their knowledge, and their elders past, present and future. 

This work has made use of data from the European Space Agency (ESA) mission
{\it Gaia} (\href{https://www.cosmos.esa.int/gaia}{\nolinkurl{cosmos.esa.int/gaia}}), processed by the {\it Gaia}
Data Processing and Analysis Consortium (DPAC, \href{https://www.cosmos.esa.int/web/gaia/dpac/consortium}{\nolinkurl{cosmos.esa.int/web/gaia/dpac/consortium}}). Funding for the DPAC
has been provided by national institutions, in particular the institutions
participating in the {\it Gaia} Multilateral Agreement. This work has made use of the \href{https://gaia-kepler.fun}{\nolinkurl{gaia-kepler.fun}} crossmatch database created by Megan Bedell.

This research made use of NASA's Astrophysics Data System; the SIMBAD database, operated at CDS, Strasbourg, France. Some of the data presented in this paper were obtained from the Mikulski Archive for Space Telescopes (MAST). STScI is operated by the Association of Universities for Research in Astronomy, Inc., under NASA contract NAS5-26555. Support for MAST for non-HST data is provided by the NASA Office of Space Science via grant NNX13AC07G and by other grants and contracts. We acknowledge the support of the Group of Eight universities and the German Academic Exchange Service through the Go8 Australia-Germany Joint Research Co-operation Scheme. 

\software{IPython  \citep{PER-GRA:2007}; SciPy \citep{jones_scipy_2001}; and Astropy, a community-developed core Python package for Astronomy \citep{astropy}.}




\bibliography{ms} 

\appendix

\startlongtable
\begin{deluxetable}{ccccccccc}
\tablecaption{The full set of underobserved and unobserved stars for which new light curves have been produced in this smear catalogue. Calibrated \gaia distances are from \citet{gaiadists}. The eclipsing binary V2083~Cyg was detected by \gaia, but a parallax could not be obtained in DR2, possibly due to binary motion.Variability classes are determined by inspection, having their usual abbreviations. EV denotes an ellipsoidal variable, and RM rotational modulation, though these two can appear similar. $\alpha^2\,\text{CVn}$ variables are chemically-peculiar stars with rotational spot modulation,and are noted separately from RM without chemical peculiarity.$\gamma\,\text{Dor} /\delta\,\text{Sct}$ denotes a $\gamma\,\text{Dor} /\delta\,\text{Sct}$ hybrid, not uncertainty.H+S denotes a `hump and spike' star.Question marks indicate uncertainty, and dashes -- that no significant variability is observed.\label{all_stars}}
\tablehead{\colhead{Object} & \colhead{KIC} & \colhead{Spectral Type} & \colhead{$Kp$} & \colhead{$G$} & \colhead{$Bp-Rp$} & \colhead{\gaia Distance} & \colhead{TRES} & \colhead{Variability}\\ \colhead{ } & \colhead{} & \colhead{(SIMBAD)} & \colhead{(mag)} & \colhead{(mag)} & \colhead{(mag)} & \colhead{(pc)} & \colhead{} & \colhead{Class}}
\startdata
\object{14 Cyg} & 7292420 & B9III & 5.490 & 5.370 & -0.055 & $194.3^{+7.0}_{-6.6}$ & -- & H+S \\
\object{BD+36 3564} & 1575741 & K5 & 8.128 & 8.041 & 1.544 & $547.1^{+11.6}_{-11.1}$ & \checkmark & RG \\
\object{BD+39 3577} & 4989821 & G5 & 8.131 & 8.090 & 1.134 & $311.7^{+2.7}_{-2.7}$ & \checkmark & RG \\
\object{BD+39 3882} & 4850372 & F5 & 8.259 & 8.159 & 0.616 & $143.3^{+0.7}_{-0.7}$ & -- & ? \\
\object{BD+42 3150} & 7091342 & K0 & 8.350 & 8.315 & 1.206 & $546.0^{+32.5}_{-29.1}$ & \checkmark & ? \\
\object{BD+42 3367} & 7447756 & M0 & 7.271 & 6.992 & 2.020 & $762.0^{+15.8}_{-15.2}$ & \checkmark & LPV \\
\object{BD+42 3393} & 6870455 & K5 & 7.664 & 7.414 & 1.952 & $929.0^{+25.9}_{-24.5}$ & \checkmark & LPV \\
\object{BD+43 3064} & 8075287 & K5 & 8.284 & 8.203 & 1.599 & $641.0^{+20.3}_{-19.1}$ & \checkmark & RG \\
\object{BD+43 3068} & 8006792 & G0 & 8.308 & 8.268 & 0.839 & $53.8^{+0.1}_{-0.1}$ & -- & -- \\
\object{BD+43 3171} & 7810954 & M0 & 8.373 & 8.178 & 1.858 & $751.5^{+17.2}_{-16.5}$ & \checkmark & LPV \\
\object{BD+43 3213} & 7747499 & K5 & 8.311 & 8.139 & 1.876 & $948.8^{+25.8}_{-24.5}$ & \checkmark & LPV \\
\object{BD+47 2825} & 10337574 & K0 & 8.251 & 8.236 & 1.329 & $485.8^{+7.3}_{-7.1}$ & -- & EB \\
\object{BD+47 2891} & 10347606 & K0 & 8.680 & 8.625 & 1.291 & $262.8^{+1.7}_{-1.6}$ & -- & RG \\
\object{BD+48 2904} & 11085556 & K0 & 8.487 & 8.439 & 1.355 & $400.9^{+5.4}_{-5.3}$ & \checkmark & RG \\
\object{BD+48 2955} & 10988024 & K2 & 7.961 & 7.899 & 1.549 & $589.4^{+11.6}_{-11.1}$ & \checkmark & RG \\
\object{HD 174020} & 7800227 & K5 & 6.753 & 6.600 & 1.754 & $433.1^{+4.2}_{-4.1}$ & \checkmark & RG \\
\object{HD 174177} & 9630812 & A2IV & 6.575 & 6.483 & 0.119 & $223.9^{+1.7}_{-1.6}$ & -- & ? \\
\object{HD 174676} & 7420037 &  & 7.481 & 7.440 & 2.434 & $993.3^{+26.7}_{-25.4}$ & \checkmark & LPV \\
\object{HD 174829} & 7339102 & K0 & 6.967 & 6.928 & 1.391 & $355.0^{+3.5}_{-3.4}$ & \checkmark & RG \\
\object{HD 175132} & 6020867 & B9IIIpSi & 6.362 & 6.242 & -0.063 & $333.3^{+5.9}_{-5.7}$ & -- & $\alpha^2\,\text{CVn}$ \\
\object{HD 175466} & 7340766 & K2 & 6.165 & 5.919 & 1.905 & $397.8^{+6.8}_{-6.6}$ & -- & LPV \\
\object{HD 175740} & 6265087 & G8III & 5.212 & 5.152 & 1.171 & $81.5^{+0.6}_{-0.6}$ & \checkmark & RG \\
\object{HD 175841} & 4989900 & A2 & 6.885 & 6.797 & 0.172 & $241.0^{+2.1}_{-2.1}$ & -- & $\gamma\,\text{Dor} /\delta\,\text{Sct}$ \\
\object{HD 175884} & 6584587 & K0 & 6.210 & 6.144 & 1.448 & $238.9^{+1.5}_{-1.4}$ & \checkmark & RG \\
\object{HD 176209} & 9327530 & A0 & 7.437 & 7.365 & 0.091 & $282.2^{+2.7}_{-2.7}$ & \checkmark & ? \\
\object{HD 176582} & 4136285 & B5V & 6.510 & 6.383 & -0.232 & $298.6^{+3.9}_{-3.8}$ & -- & $\alpha^2\,\text{CVn}$ \\
\object{HD 176626} & 7943968 & A2V & 6.933 & 6.841 & 0.035 & $224.8^{+1.8}_{-1.7}$ & -- & RM \\
\object{HD 176894} & 6267965 & F0 & 7.700 & 7.610 & 0.530 & $82.8^{+0.2}_{-0.2}$ & -- & $\gamma\,\text{Dor}$ \\
\object{HD 177697} & 4994443 & K5 & 7.300 & 6.764 & 2.338 & $472.0^{+5.4}_{-5.3}$ & -- & RG \\
\object{HD 177781} & 2970780 & G5 & 7.744 & 7.701 & 1.024 & $296.2^{+2.6}_{-2.5}$ & -- & $\gamma\,\text{Dor} /\delta\,\text{Sct}$ \\
\object{HD 178090} & 6675338 & K5 & 6.758 & 6.549 & 1.892 & $583.0^{+8.5}_{-8.3}$ & -- & LPV \\
\object{HD 178797} & 10064283 & K0 & 7.312 & 7.249 & 1.478 & $406.1^{+4.8}_{-4.7}$ & \checkmark & RG \\
\object{HD 178910} & 11288450 & K2 & 7.864 & 7.848 & 1.346 & $291.3^{+2.4}_{-2.4}$ & \checkmark & RG \\
\object{HD 179394} & 7105221 & B8 & 7.575 & 7.475 & -0.100 & $476.2^{+12.2}_{-11.6}$ & \checkmark & -- \\
\object{HD 179395} & 6593264 & B9 & 7.168 & 7.070 & 0.067 & $233.9^{+1.7}_{-1.7}$ & -- & $\alpha^2\,\text{CVn}$ \\
\object{HD 179396} & 3838362 & K2 & 8.001 & 7.970 & 1.244 & $321.2^{+2.7}_{-2.6}$ & \checkmark & RG \\
\object{HD 179959} & 10265370 & K0 & 6.280 & 6.258 & 1.168 & $499.2^{+7.2}_{-7.0}$ & \checkmark & RG \\
\object{HD 180312} & 4551179 & K0II & 7.970 & 7.834 & 1.162 & $290.5^{+2.4}_{-2.4}$ & \checkmark & RG \\
\object{HD 180475} & 11656042 & K2 & 7.664 & 7.595 & 1.489 & $546.1^{+8.0}_{-7.8}$ & \checkmark & RG \\
\object{HD 180658} & 6195870 & K0 & 7.932 & 7.871 & 1.256 & $282.2^{+2.3}_{-2.3}$ & \checkmark & RG \\
\object{HD 180682} & 5177450 & K0 & 6.617 & 6.532 & 1.486 & $295.8^{+2.5}_{-2.5}$ & \checkmark & LPV \\
\object{HD 181022} & 3946721 & K5 & 6.496 & 6.248 & 1.892 & $317.7^{+2.7}_{-2.7}$ & \checkmark & LPV \\
\object{HD 181069} & 4049174 & K1III & 6.279 & 6.264 & 1.237 & $144.2^{+0.6}_{-0.6}$ & \checkmark & RG \\
\object{HD 181097} & 4149233 & K0 & 7.920 & 7.848 & 1.434 & $434.3^{+6.2}_{-6.0}$ & \checkmark & RG \\
\object{HD 181328} & 12456737 & M1 & 7.182 & 6.614 & 2.334 & $353.9^{+3.3}_{-3.3}$ & \checkmark & LPV \\
\object{HD 181521} & 5180075 & A0 & 6.939 & 6.852 & 0.059 & $217.8^{+3.4}_{-3.3}$ & -- & $\gamma\,\text{Dor} /\delta\,\text{Sct}$ \\
\object{HD 181596} & 11910615 & K5III & 7.050 & 6.863 & 1.841 & $591.1^{+8.1}_{-7.8}$ & \checkmark & RG \\
\object{HD 181597} & 11555267 & K1III & 6.040 & 5.985 & 1.283 & $135.8^{+0.3}_{-0.3}$ & \checkmark & RG \\
\object{HD 181681} & 5092997 & K4III & 6.864 & 6.696 & 1.798 & $585.0^{+9.1}_{-8.9}$ & \checkmark & RG \\
\object{HD 181778} & 7816792 & K0 & 7.545 & 7.514 & 1.315 & $374.5^{+3.4}_{-3.4}$ & \checkmark & RG \\
\object{HD 181878} & 4830109 & G5 & 6.698 & 6.587 & 1.003 & $259.5^{+1.8}_{-1.8}$ & \checkmark & RG \\
\object{HD 181880} & 3337423 & K & 7.982 & 7.940 & 1.498 & $541.2^{+10.1}_{-9.7}$ & \checkmark & RG \\
\object{HD 182354} & 2156801 & K0 & 6.320 & 6.291 & 1.253 & $228.9^{+1.7}_{-1.7}$ & \checkmark & RG \\
\object{HD 182531} & 11188366 & K5 & 7.955 & 7.859 & 1.502 & $599.3^{+9.2}_{-8.9}$ & \checkmark & RG \\
\object{HD 182692} & 10728753 & K0 & 7.310 & 7.247 & 1.227 & $226.6^{+1.3}_{-1.3}$ & \checkmark & RG \\
\object{HD 182694} & 7680115 & G7IIIa & 5.722 & 5.598 & 1.061 & $133.1^{+0.7}_{-0.7}$ & \checkmark & RG \\
\object{HD 182737} & 1572070 & A0 & 7.820 & 7.758 & 0.421 & $460.3^{+6.7}_{-6.5}$ & -- & RM \\
\object{HD 183124} & 8752618 & G8II & 6.441 & 6.395 & 1.176 & $160.7^{+0.8}_{-0.8}$ & \checkmark & RG \\
\object{HD 183203} & 12208512 & K5 & 6.928 & 6.530 & 2.116 & $476.9^{+5.9}_{-5.8}$ & \checkmark & LPV \\
\object{HD 183362} & 2715115 & B3Ve & 6.394 & 6.208 & -0.041 & $571.1^{+18.2}_{-17.2}$ & -- & $\gamma\,\text{Dor}$,\,H+S \\
\object{HD 183383} & 6777469 & B9 & 7.640 & 7.537 & 0.081 & $357.1^{+5.5}_{-5.3}$ & -- & ? \\
\object{HD 184147} & 9651435 & B9IV & 7.251 & 7.145 & -0.037 & $175.5^{+2.6}_{-2.5}$ & -- & ? \\
\object{HD 184215} & 11031549 & B8 & 7.321 & 7.189 & -0.135 & $361.2^{+6.4}_{-6.1}$ & -- & SPB \\
\object{HD 184483} & 7756961 & M5 & 7.246 & 6.719 & 2.337 & $492.9^{+5.5}_{-5.4}$ & \checkmark & LPV \\
\object{HD 184565} & 6047321 & K0 & 7.972 & 7.943 & 1.024 & $380.9^{+4.3}_{-4.2}$ & -- & LPV \\
\object{HD 184787} & 6528001 & A0V & 6.757 & 6.658 & -0.003 & $139.6^{+1.1}_{-1.1}$ & \checkmark & H+S \\
\object{HD 184788} & 6129225 & B9 & 7.249 & 7.143 & -0.055 & $226.5^{+2.4}_{-2.3}$ & -- & RM \\
\object{HD 184875} & 6954647 & A2V & 5.403 & 5.279 & 0.107 & $172.6^{+3.3}_{-3.2}$ & -- & $\gamma\,\text{Dor}$ \\
\object{HD 185117} & 9094435 & K5 & 7.696 & 7.472 & 1.921 & $817.7^{+14.8}_{-14.3}$ & -- & LPV \\
\object{HD 185286} & 7966681 & K5 & 6.151 & 6.055 & 1.645 & $263.5^{+3.9}_{-3.8}$ & \checkmark & RG \\
\object{HD 185351} & 8566020 & G8.5IIIbFe-0.5 & 5.034 & 4.882 & 1.091 & $41.2^{+0.1}_{-0.1}$ & \checkmark & RG \\
\object{HD 185397} & 3455268 & A5 & 6.953 & 6.855 & 0.421 & $180.0^{+1.0}_{-1.0}$ & -- & $\delta\,\text{Sct}$ \\
\object{HD 185524} & 8960196 & K2 & 8.022 & 7.953 & 1.368 & $753.4^{+15.9}_{-15.2}$ & \checkmark & LPV \\
\object{HD 186121} & 7456762 & M3III & 5.773 & 5.176 & 2.250 & $475.2^{+35.1}_{-30.7}$ & \checkmark & LPV \\
\object{HD 186155} & 9163520 & F5II-III & 5.055 & 4.923 & 0.529 & $50.6^{+0.4}_{-0.4}$ & -- & H+S \\
\object{HD 186255} & 4937492 & A3 & 6.966 & 6.862 & 0.252 & $254.5^{+4.1}_{-4.0}$ & -- & $\delta\,\text{Sct}$ \\
\object{HD 186727} & 12316020 & M0 & 7.499 & 6.917 & 2.388 & $581.7^{+9.2}_{-8.9}$ & \checkmark & LPV \\
\object{HD 186994} & 8766240 & B0III & 7.585 & 7.451 & -0.185 & $1866.1^{+138.1}_{-120.6}$ & -- & EB \\
\object{HD 187217} & 11824273 & K0 & 6.399 & 6.345 & 1.273 & $243.2^{+1.8}_{-1.8}$ & \checkmark & RG \\
\object{HD 187277} & 6967644 & A0 & 7.579 & 7.464 & 0.282 & $96.9^{+0.4}_{-0.4}$ & -- & -- \\
\object{HD 187372} & 10679281 & M1III & 5.672 & 5.313 & 2.047 & $306.4^{+10.3}_{-9.6}$ & \checkmark & LPV \\
\object{HD 188252} & 10683303 & B2III & 6.007 & 5.864 & -0.276 & $1000.6^{+82.6}_{-71.1}$ & -- & SPB \\
\object{HD 188537} & 9110718 & K0 & 7.382 & 7.324 & 1.345 & $629.9^{+11.4}_{-11.0}$ & \checkmark & RG \\
\object{HD 188629} & 8710324 & K5 & 7.743 & 7.546 & 1.888 & $651.0^{+12.0}_{-11.6}$ & \checkmark & LPV \\
\object{HD 188875} & 5041881 & K2 & 6.164 & 6.091 & 1.584 & $683.8^{+12.4}_{-11.9}$ & \checkmark & RG \\
\object{HD 189013} & 10096499 & A2 & 6.922 & 6.840 & 0.225 & $188.8^{+6.4}_{-6.0}$ & -- & $\gamma\,\text{Dor}$ \\
\object{HD 189178} & 5219588 & B5V & 5.552 & 5.410 & -0.106 & $347.3^{+13.0}_{-12.1}$ & -- & SPB,\,H+S \\
\object{HD 189636A} & 10298067 &  & 8.025 & 8.118 & 1.211 & $384.7^{+6.0}_{-5.8}$ & -- & ? \\
\object{HD 189636B} & 10298061 &  & 8.107 & 8.024 & 1.316 & $376.4^{+4.9}_{-4.7}$ & -- & ? \\
\object{HD 189684} & 9305008 & A5III & 5.982 & 5.881 & 0.246 & $125.2^{+6.2}_{-5.7}$ & -- & EV \\
\object{HD 189750} & 8521828 & K0 & 8.052 & 8.061 & 1.207 & $327.0^{+3.0}_{-2.9}$ & \checkmark & ? \\
\object{HD 190149} & 8262528 & M0II-III & 6.488 & 6.171 & 2.031 & $409.4^{+3.8}_{-3.7}$ & \checkmark & LPV \\
\object{HD 226754} & 6234579 & K2 & 7.829 & 7.702 & 1.652 & $391.8^{+6.1}_{-5.9}$ & \checkmark & RG \\
\object{V2079 Cyg} & 8818020 & B8V & 7.174 & 7.034 & -0.221 & $321.5^{+3.7}_{-3.6}$ & -- & $\alpha^2\,\text{CVn}$ \\
\object{V2083 Cyg} & 10342012 & A3 & 6.902 & 6.813 & 0.351 & -- & -- & EB \\
\object{V380 Cyg} & 5385723 & B1.1III+B2.5/3V: & 5.771 & 5.632 & -0.062 & $1044.7^{+116.6}_{-95.6}$ & -- & EB \\
\object{V398 Lyr} & 4042516 & M3 & 7.024 & 5.403 & 3.406 & $494.7^{+34.9}_{-30.6}$ & \checkmark & RG \\
\object{V543 Lyr} & 5429169 & B3V & 6.299 & 6.160 & -0.217 & $345.1^{+5.6}_{-5.4}$ & -- & SPB \\
\object{V546 Lyr} & 6267345 & M3III & 7.385 & 6.784 & 2.443 & $587.8^{+13.1}_{-12.6}$ & \checkmark & LPV \\
\object{V547 Lyr} & 5429948 & M4-IIIa & 6.199 & 5.228 & 2.725 & $288.9^{+13.1}_{-12.0}$ & \checkmark & LPV \\
\object{V554 Lyr} & 5001462 &  & 8.179 & 8.092 & -0.129 & $335.7^{+4.6}_{-4.5}$ & -- & $\alpha^2\,\text{CVn}$ \\
\object{V819 Cyg} & 10618721 & B0.5IIIn & 6.381 & 6.243 & -0.160 & $1114.0^{+70.9}_{-63.0}$ & -- & SPB
\enddata
\end{deluxetable}





\end{document}